%% file: main.tex
\documentclass{esagnc}

\graphicspath{{figures/}}

\begin{document}

\papertitle{TRAJECTORY OPTIMISATION OF A SWARM ORBITING 67P/CHURYUMOV-GERASIMENKO MAXIMISING GRAVITATIONAL SIGNAL}

\mainauthor{R. Maråk}     
\author[(1)]{Rasmus Maråk}
\author[(2)]{Emmanuel Blazquez}
\author[(2, 3)]{Pablo Gómez}
\affil[(1)]{\textit{Optimization and Systems Theory, Department of Mathematics, KTH Royal Institute of Technology, Lindstedtsvägen 25, 114 28 Stockholm, Sweden, rasmusmarak@gmail.com}}
\affil[(2)]{\textit{Advanced Concepts Team, European Space Agency, European Space Research and Technology Centre (ESTEC), Keplerlaan 1, 2201 AZ Noordwijk, The Netherlands, \{Pablo.Gomez, Emmanuel.Blazquez}@esa.int\}}
\affil[(3)]{\textit{AI Sweden, Lindholmspiren 11, 417 56 Göteborg, Sweden}}

\paperabstract{
Proper modelling of the gravitational fields of irregularly shaped asteroids and comets is an essential yet challenging part of any spacecraft visit and flyby to these bodies. Accurate density representations provide crucial information for proximity missions which rely heavily on it to design safe and efficient trajectories. This work explores using a spacecraft swarm to maximise the measured gravitational signal in a hypothetical mission around the comet 67P/Churyumov-Gerasimenko. Spacecraft trajectories are simultaneously propagated with an evolutionary optimisation approach to maximise overall signal return. The propagation is based on an open-source polyhedral gravity model using a detailed mesh of 67P and considers the comet’s sidereal rotation. We compare performance on a mission scenario using one and four spacecraft. The results show that the swarm achieved almost twice the single spacecraft coverage over a fixed mission duration. However, optimising for a single spacecraft results in a more effective trajectory. Overall, this work serves as a testbed for efficiently designing a set of trajectories in this complex gravitational environment balancing measured signals and risks in a swarm scenario. 
The codebase and results are publicly available at \url{https://github.com/rasmusmarak/TOSS}.
}

\input{Sections/Introduction}
\input{Sections/RelatedWork}

\input{Sections/Methods}
\input{Sections/Results}
\input{Sections/Discussion}
\input{Sections/Conclusion}
\input{Sections/Acknowledgements}

\printbibliography
\end{document}

%% file: Sections/Introduction.tex
\section{Introduction}\label{section 1}
\subsection{Background}\label{section 1.1}
Recently, there has been an increased interest in exploring smaller celestial bodies such as asteroids and comets. It is believed that most of these bodies are remnants from the formation of our solar system and, thus, act as geological time capsules. Studying their gravitation field can give valuable clues on their elemental composition and, consequently, a critical insight into the formation and evolution of our solar system \cite{Nucleus67P, Cui2014, Oguri2021}. Another interest in exploring these bodies links to planetary defence, which stems from the risk of hazardous collisions provided their relative proximity to Earth. For these reasons, space agencies worldwide have conducted various missions to these bodies for the last two decades, including targets of observation, landing and collecting samples. Prominent examples include the Rosetta mission by ESA landing on the comet 67P/Churyumov-Gerasimenko \cite{Rosetta}, JAXA's missions Hayabusa I and II to the asteroids 25143 Itokawa and 162173 Ryugu \cite{Fujiwara2006, SAIKI2021}, the Origins, Spectral Interpretation, Resource Identification Security-Regolith Explorer (known as OSIRIS-REx) to the asteroid 101955 Bennu \cite{Nolan2013} as well as NASA's Double Asteroid Redirection Test (DART) \cite{DART} and the upcoming complementary mission Hera by ESA \cite{HERA1, HERA2}, both to the Didymos asteroid system.

For prospective missions to these small bodies, it is essential to find efficient and safe trajectories. However, designing such missions is rather challenging for several reasons. Compared to planets and moons, asteroids and comets are much smaller, rendering a weaker gravitational field, irregular shape and often heterogeneous mass density \cite{Broschart2005}. Altogether, these factors result in complex dynamics where ideal Keplerian motion becomes unstable, and the orbital trajectories are oftentimes non-periodic, thus resulting in either escape trajectories or collisions with the body itself \cite{Furfaro2021}. With these prospects in mind, it becomes evident that proper modelling of the gravitational fields makes up an essential part of any spacecraft proximity missions to any smaller, irregularly shaped celestial body.

Traditionally, most small body missions have used a single spacecraft to carry out its designated scientific measurements \cite{Fujiwara2006, Nolan2013, Rosetta}. In recent years, however, more work has been seen utilising spacecraft swarms to estimate the gravitational field around small bodies \cite{VillaJ2022, Stacey2022, stacey2018autonomous}. Another prominent example is the upcoming mission Hera, by ESA. With significant advances in autonomous control and miniaturisation of embedded systems, the Hera mission will use two 6U CubeSats to visit the Didymos asteroid system and measure the asteroid’s size, shape, and composition as well as the impact of DART \cite{HERA1, HERA2, DART}. The benefits of utilising multiple spacecraft for these purposes are many. For instance, it allows for more efficient and cost-effective missions where smaller spacecraft can be launched more frequently and at a lower cost than larger spacecraft. Utilising a spacecraft swarm can also be beneficial depending on the specific mission scenario as it enables multiple simultaneous observations and reduces the cost of a potential collision with the body \cite{CastilloRogez2019}.


In this paper, we propose a method based on a direct heuristic approach for optimising the trajectories of a spacecraft swarm employing an evolutionary optimisation scheme to determine the initial state and control variables over a given mission duration. The equations of motion are solved with an adaptive step numerical integrator and then re-sampled for a fixed time step using the corresponding dense output interpolation constants. The performance of each trajectory is evaluated using a set of reward and penalty functions relating to the coverage of a predefined region where the gravitational signal is maximal, collision risks and the distance to the region of interest. We rely on a series of physical models, a polyhedral gravity model, as the available shape model to accurately represent the spatial variation of the gravity field around the small rotating body. Furthermore, the method is applied to a prospective mission to the highly irregular and rotating comet 67P/Churyumov-Gerasimenko.

The contribution of this work is primarily the formulation of a nonlinear optimisation problem that considers a constellation of spacecraft to optimise the measured gravitational signal around a smaller body with challenging dynamics. Although the proposed method will be particularly suited for finding feasible trajectories for the training and robustness evaluation of GeodesyNets \cite{GeodesyNets2, GeodesyNetsBenchmark}, it is generally relevant for a broader range of mission scenarios. In detail, the corresponding code for solving this problem is an open-source modular codebase\footnote{Codebase for this work: \href{https://github.com/rasmusmarak/TOSS}{https://github.com/rasmusmarak/TOSS}}.

This paper is organised as follows. First, we begin by briefly reviewing previous work on trajectory optimisation near smaller celestial bodies, particularly focusing on swarm optimisation and evolutionary approaches for these tasks. Further, we describe the proposed method for finding feasible trajectories that maximise the measured gravitational signal. We then evaluate the approach by applying it to a high-fidelity polyhedral model of the comet 67P/Churyumov-Gerasimenko and review several performance measures, such as quality of measured signal over time, convergence rate and average distance to the region of interest.

%% file: Sections/RelatedWork.tex
\section{Related work}\label{section 2}
\subsection{Modelling Gravitational Fields}\label{section 2.1}
When modelling the gravitational field, the literature often presents three models: spherical harmonics, mascon models and polyhedral gravity models. The former class, spherical harmonics, is a multi-pole expansion method where the accelerations can be computed using the spatial derivatives of the potential harmonics \cite{Furfaro2021}. However, the convergence of the model is generally limited to the points outside the Brillouin sphere - the sphere co-centred with the expansion and precisely covering the complete body - which renders the model unsuitable for irregularly shaped celestial bodies. Although the two latter classes are more suited for these bodies, their performance is generally constrained by their corresponding shape model assumptions. For instance, mascon models approach the problem by dividing the body into smaller mass concentrations called mascons, consequently limiting the performance by the detail of the discretisation. For the latter class, the polyhedral gravity model, it is assumed that the body of interest can be approximated by a polyhedron of constant mass density, enabling a more accurate description of the gravitational field \cite{Werner1997}. However, the model generally suffers from computationally expensive tasks depending on its resolution and the fact that most of these bodies tend to have mass-density heterogeneity.

Recent advances in machine learning and neural networks have increased interest in developing new data-driven models that aim to provide more accurate models of these complex celestial bodies. One such model is the neural density field, a versatile tool that can accurately describe the density distribution of a body’s mass and internal and external shape with few prior requirements. GeodesyNets is one approach that attempts to solve the gravity inversion problem by learning a three-dimensional, differentiable function representing the body density, referred to as the neural density field \cite{GeodesyNets1}. By training on either real or synthetic ground truth data, the body’s shape and other geodetic properties can quickly be recovered utilising that its integration leads to a gravitational field model. This representation has several advantages as it requires no prior information on the body, converges inside the Brillouin sphere, and is extensible even to heterogeneous density distributions. Thus avoiding most issues in the three models mentioned above \cite{GeodesyNets1}. Previous work includes training the model on synthetic ground truth data without any previous shape models and training on data recovered from the OSIRIS-REx mission visiting the asteroid 101955 Bennu and an available shape model \cite{OsirisRex1, GeodesyNets2}. Although training on synthetic data showed comparable performance with a polyhedral gravity model assuming mass density homogeneity \cite{GeodesyNets1}, the last test did not show any significant improvement in the fidelity of the modelled gravitational field, most likely due to uncertainties of non-gravitational forces in the propagation of trajectories. Therefore, it is of great interest to generate more realistic trajectories based on a polyhedral gravity model to study geodesyNets' performance further.

\subsection{Trajectory Optimisation of Spacecraft Swarms}\label{section 2.2}


In the literature related to small body exploration, utilising spacecraft swarms has shown promising results in terms of scientific returns compared to more traditional monolithic approaches \cite{Rossi2022}. With the prospective advantages of constellations, numerous papers have proposed efficient methods for designing trajectories around smaller bodies to obtain a high-accuracy mapping of body attributes and efficient spacecraft navigation. In terms of direct approaches, examples include mixed-integer linear programming schemes with gradient-based global search processes \cite{Rossi2022}, sampling-based model predictive optimisation frameworks \cite{Capolupo2017} and two-staged optimisation schemes including a nonlinear optimisation process for trajectory planning and a genetic multi-spacecraft travelling salesman problem \cite{Satpute2021}. Although these articles have used widely different approaches, a common feature is that swarm missions tend to require the simultaneous optimisation of several competing objectives such as viewing angles, keeping in line of sight for communication purposes, minimising fuel consumption, maximising some defined measure of coverage and avoiding impacts. In order to satisfy these objectives, the optimisation space expands, creating an even greater demand for robust and efficient search strategies of the feasible space. For these types of problems, population-based optimisation schemes are well suited. With similar motivation, Pearl, J.M et al \cite{pearl2019curvilinear} proposes an alternative approach using a curvilinear surface-based gravity model and differential evolution for optimising the initial state of a single-spacecraft trajectory. The performance of the candidate solutions is then compared by integrating the system dynamics and computing the fitness along the state propagation.

With motivation in the efficiency and robustness of evolutionary approaches, this work considers evolutionary methods for determining optimal trajectories corresponding to a spacecraft swarm orbiting a small celestial body with challenging dynamics. Specifically, we consider the comet 67P/Churyumov-Gerasimenko given its highly irregular shape and rotating motion. Furthermore, by modelling the dynamics around the body using a polyhedral gravity model, the resulting trajectories will be more feasible to study the performance of geodesyNets more realistically.

%% file: Sections/Methods.tex
\section{Methods}\label{section 3}

\subsection{Problem Formulation}\label{section 3.1}
In most cases, a spacecraft trajectory optimisation problem can effectively be represented as an optimal control problem. The goal is typically to determine a control strategy that minimises some performance measures for a state space model governed by continuous dynamics. In general, the trajectory optimisation problem can be formulated as\\
\begin{equation}
    \begin{aligned}
        \min_{\boldsymbol{x}, \boldsymbol{u}} \quad & \Phi(t_f, \boldsymbol{x_f}) + \int^{t_f}_{t_0} \mathcal{L}(\boldsymbol{x}(t), \boldsymbol{u}(t), t) dt \\
        \textrm{s.t.} \quad & \dot{\boldsymbol{x}} = f(\boldsymbol{x},t) + \boldsymbol{u}(t)\\
          & \Psi(t_0,t_f,\boldsymbol{x_0},\boldsymbol{x_f}) \leq 0 \\
          & \boldsymbol{u} \in \mathcal{U}(\boldsymbol{x},t)
    \end{aligned}
    \label{eq:OptimalControlProblem}
\end{equation}\\
where the state vector $\boldsymbol{x}(t)$ represents the orbital state of the spacecraft, $f(\boldsymbol{x},t)$ the external forces acting on the spacecraft in motion, $\boldsymbol{u}(t)$ the control forces acting on the spacecraft by its propulsion, $\Psi$ the boundary constraints on the initial and final states, $\Phi(t_f, \boldsymbol{x_f})$ the terminal cost and $\mathcal{L}(\boldsymbol{x}(t), \boldsymbol{u}(t), t)$ a running cost function. Furthermore, it should be noted that the set of admissible controls, $\mathcal{U}(\boldsymbol{x},t)$, is presented with a state dependency which is generally uncommon for optimal control problems but relevant here given how the use of manoeuvres are reasonably related to the spacecraft's position and motion. Another characteristic of Equation \ref{eq:OptimalControlProblem} is the complicated nature of the differential system $\dot{\boldsymbol{x}}$, which tends to induce several locally optimal solutions making more straightforward numerical methods unsuitable while also requiring more robust and accurate global optimisation techniques to cover the search space efficiently. In this subsection, we aim to reduce the complexity of the optimal control problem by reformulating it into a more simple optimisation problem suitable for evolutionary methods. For this, we will introduce the relevant equations of motion and cost functions for defining a set of trajectories that maximises the measured gravitational signal. We will first consider the general problem for one spacecraft for simplicity and then introduce a swarm.

\subsubsection{Equations of Motion}\label{section 3.1.1}
To model the equations of motion for a single spacecraft, we introduce a state-space representation where $\boldsymbol{x}(t)$ defines the orbital state in cartesian coordinates through its position $\boldsymbol{r}(t) \in \mathbb{R}^3$ and velocity $\boldsymbol{v}(t) \in \mathbb{R}^3$. The differential system in the vicinity of the body can be formulated as\\
\begin{equation}
    \dot{\boldsymbol{x}} = 
    \begin{bmatrix}
        \dot{\boldsymbol{r}}(t)\\
        \dot{\boldsymbol{v}}(t)
    \end{bmatrix}
    =
    \begin{bmatrix}
        \boldsymbol{v}(t)\\
        \boldsymbol{g}(\boldsymbol{r}(t)) + \Gamma(t) u(t)
    \end{bmatrix}
    \label{eq:FirstEqOfMotion}
\end{equation}\\
with the initial state as $\boldsymbol{x}(t_0)$, unit control vector $u(t)$ and the magnitude of the control as $\Gamma(t)$. In Equation \ref{eq:FirstEqOfMotion}, $\boldsymbol{g}(\boldsymbol{r}(t))$ is the gravitational force acting on the spacecraft by the comet or asteroid. In this work, the acceleration at any given position can be computed from the polyhedral gravity model as presented by Tsoulis, D \cite{Tsoulis2012, Tsoulis2021} and further implemented by Schuhmacher, J. \cite{PolyhedralModelJonas}\footnote{Polyhedral gravity model implementation: \href{https://github.com/esa/polyhedral-gravity-model}{https://github.com/esa/polyhedral-gravity-model} (Accessed: 30-01-2023)}. In reality, however, it is not uncommon for comets and asteroids to spin, where the now rotating gravitational field consequently affects the acceleration of the orbiting spacecraft. Therefore, given that irregular gravitational fields and body-relative geometric constraints are more realistically expressed considering the body's rotational motion, the gravitational force acting on the spacecraft will be computed in the fixed-body reference frame. For this purpose, we introduce a rotation quaternion $\boldsymbol{q} =(q_0,q_1,q_2,q_3)$ around the small-body principal axis $\boldsymbol{\omega}$, defined by the its declination and right ascension.


Furthermore, the orbit control vector $\boldsymbol{u}(t)$ is modelled by impulsive manoeuvres representing an instantaneous change in velocity in the state vector applied at discrete times $\mathcal{T}_{C} = \{ t_{i}: t_{i} < t_{i+1}, i=0,1, \dots, I \}$. Assuming that the body is rotating with constant angular velocity, that is, excluding the effects of outgassing and similar perturbations, the equations of motion presented in Equation \ref{eq:FirstEqOfMotion} can be further refined as\\
\begin{equation}
    \dot{\boldsymbol{x}} = 
    \begin{bmatrix}
        \dot{\boldsymbol{r}}_t\\
        \dot{\boldsymbol{v}}_t
    \end{bmatrix}
    =
    \begin{bmatrix}
        \boldsymbol{v}_t\\
        \boldsymbol{a}(R_{\boldsymbol{q}}(\boldsymbol{w}, \nu, t)\boldsymbol{r}_t) + \boldsymbol{u}_{t|t\in \mathcal{T}_{C}}
    \end{bmatrix}
    \label{eq:RefinedEqOfMotion}
\end{equation}\\
To obtain a stable solution, the differential system presented in Equation \ref{eq:RefinedEqOfMotion} must be accurately integrated for some domain-feasible initial state $\boldsymbol{x_0} = [\boldsymbol{r}(t_0),\boldsymbol{v}(t_{0})]^T$ and a small enough time step $\Delta t$. With previous studies showing promising results for applying an eighth-order Runge-Kutta scheme with a seventh-order embedded method for adaptive step size control for orbit propagation \cite{Atallah2019, Tsoulis2016}, this work uses the Dormand-Prince 8(7)-13M numerical integration scheme \cite{Prince1981}\footnote{Integrator provided by DEsolver library: \href{https://github.com/Microno95/desolver}{https://github.com/Microno95/desolver} (Accessed: 23-02-2023)}. 

\subsubsection{Fitness Function}\label{section 3.1.2}
\begin{figure}[h]
     \centering
     \begin{subfigure}[b]{0.4\textwidth}
         \centering
         \includegraphics[width=\textwidth]{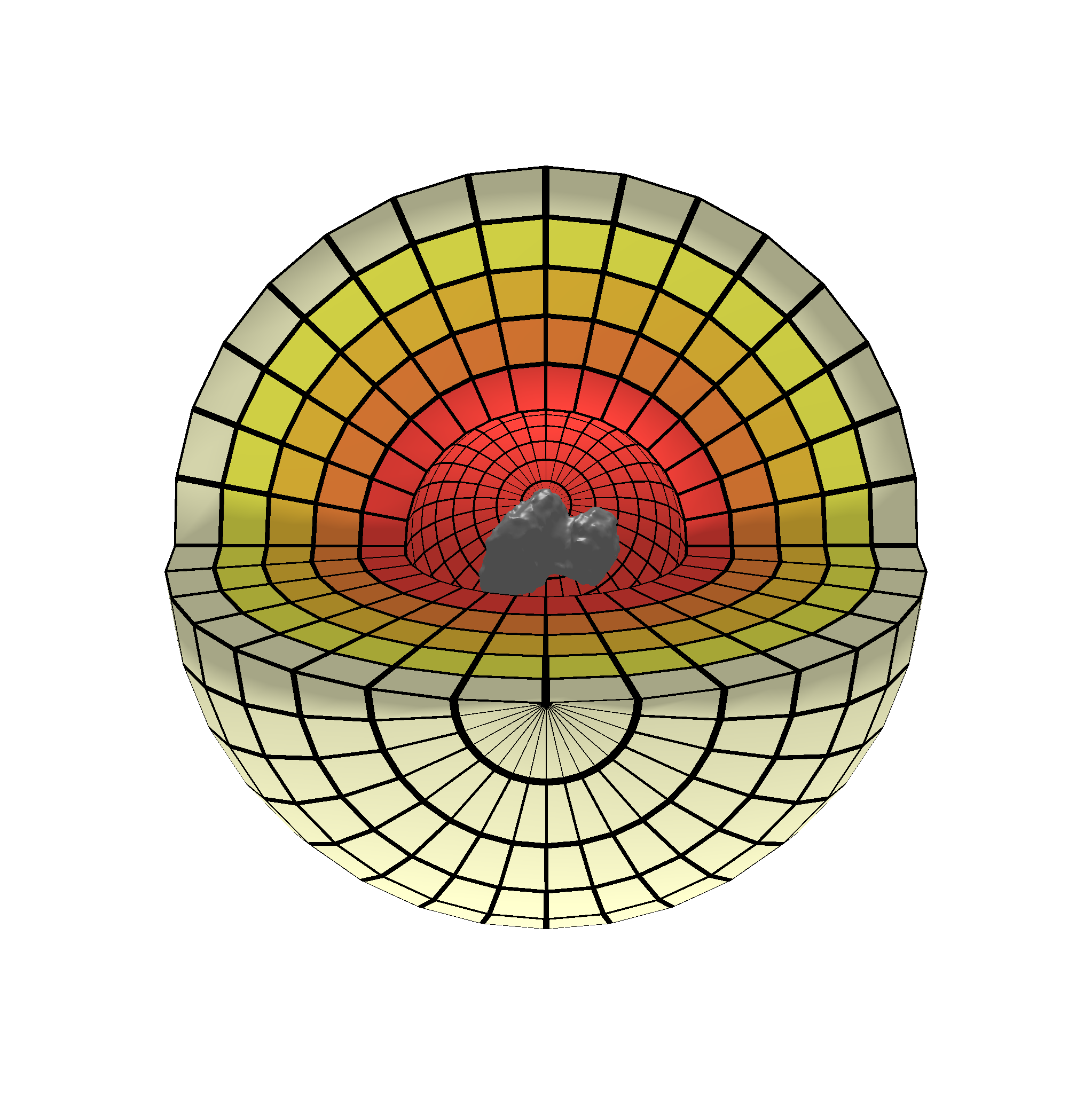}
         \caption{Spherical tensor grid around the comet.}
         \label{fig:SphericalGrid}
     \end{subfigure}
     \begin{subfigure}[b]{0.4\textwidth}
         \centering
         \includegraphics[width=0.8\textwidth]{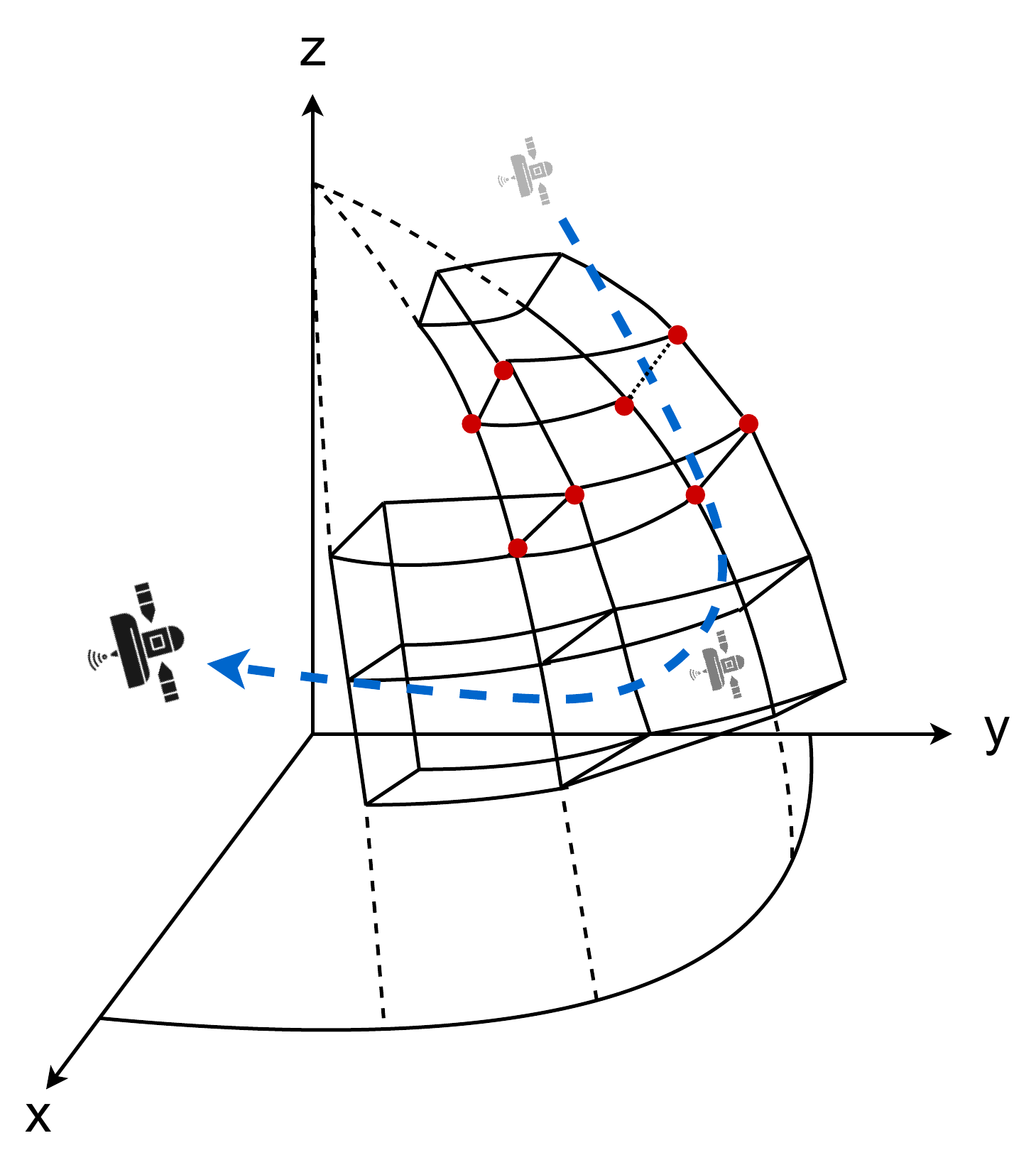}
         \caption{Trajectory crossing tesseroids on the grid.}
         \label{fig:tesseroid}
     \end{subfigure}
     \caption{Discretised gravitational field around the comet depicting a) a spherical tensor grid approximation}{with one quarter of the grid removed for visibility and b) a set of tesseroids on the spherical grid being visited by the trajectory.}
\end{figure}

To model the maximisation of the measured gravitation signal, we introduce a fitness function representing the spacecraft's coverage of the surrounding environment along the trajectory. Realistically, the measurement is a continuous process most accurately represented by integrating a corresponding continuous function. However, performing a detailed integration process in addition to integrating the dynamical equations is likely to be computationally expensive. Another strategy would be to define a discrete approximation of some bounded space around the body of interest. We introduce two new spherical constraints depicting a non-convex feasible region of positions to limit the search space for the decision vector. The constraints are in the form of bounding spheres, where the inner sphere $\mathcal{S}_I$ represents a region with greater risk for colliding with the body and is defined by a safety radius $r_{\mathcal{I}}$, and the outer sphere $\mathcal{S}_O$, with radius $r_{\mathcal{O}} > r_{\mathcal{I}}$, represents the space that is of interest for measuring the gravitational signal. The feasible region can therefore be defined by the set $\mathcal{S}_f = \{ (r, \theta, \varphi) \text{ }|\text{ } r \in [r_{\mathcal{I}}, r_{\mathcal{O}}], \theta \in [-\frac{\pi}{2},\frac{\pi}{2}], \varphi \in [-\pi, \pi] \}$. Furthermore, a discrete approximation of $\mathcal{S}_{f}$ can consequently be represented by a spherical tensor grid $\mathcal{G}$ as seen in Figure \ref{fig:SphericalGrid}, consisting of several evenly spread points $\boldsymbol{p} \in \mathcal{G}$, such that $\mathcal{G} \subset \mathcal{S}_{f}$. As a results, each set of eight adjacent points on $\mathcal{G}$ defines a tesseroid bounded by six differential surfaces as depicted in Figure \ref{fig:tesseroid}.
\\
In order to assure stability in the simulation of coverage and produce accurate results, the spacing between each point on the spherical tensor must be defined such that each tesseroid satisfies the Courant-Friedrichs-Lewy condition \cite{Gnedin2018}. In other words, we ensure that a spacecraft can not cross multiple tesseroids in one time step. This is done by considering a large enough magnitude for the spacecraft's velocity, $v_{max}$, in the inertial reference frame and a fixed measurement period corresponding to a fixed time step $\Delta t$, which together defines the length interval $\Delta L$ for the smallest possible regions, that is, the innermost set of tesseroids. Conversely, the Courant number equals a value of one. Then, using the interpolation constants provided by the dense output from the adaptive step integrator, the trajectory is re-sampled for the new set of times $\mathcal{T}_{T} = \{t_{0}\leq t \leq t_{f}: t_{n+1}-t_{n} = \Delta t, n=0,\dots,f \}$. Consequently, ensuring that the discretization of the problem remain proportional to the grid spacing. With appropriately chosen values on $\Delta t$ and $v_{max}$, the coverage, $V_c$, is the ratio of grid cells visited by the spacecraft along its trajectory. In order to assure that the coverage behaves proportionally to $1/r$, each point on the grid $p \in \mathcal{G}$ is assigned a weight defined by its radial component, $w_p = 1/r_p$. In addition, since the spherical tensor grid represents a discrete gravitational field approximation, it should be adjusted for the comet's sidereal rotation. However, to avoid the computational cost of rotating the points defined on the grid, the coverage will consider evaluating the trajectory adjusted for the body-fixed reference frame instead.


With the complex dynamics and low mass of the bodies considered in this work, and depending on the radius of the outer bounding sphere, it is not unrealistic that the most efficient solutions include positions outside the region of interest, mainly due to hyperbolic and highly non-periodic trajectories. To allow such positions, the non-convex constraint $r(t) \in \mathcal{S}_{f}$ is relaxed by introducing two smooth penalty functions that become active whenever the constraint is violated. The former function, $P_{I}$, relates the risk of colliding with the body by the closest point on the trajectory, defining a penalty based on the normalised squared distance to the safety radius.The latter function, $P_{O}$, returns a normalised penalty dependent on the average distance to the feasible region for each position defined outside the outer bounding sphere. In order to avoid additional ruggedness in the objective when comparing the penalties with the coverage, we introduce appropriately chosen penalty scaling factors $\delta_{p_{\mathcal{I}}}$ and $\delta_{p_{\mathcal{O}}}$. Finally, the complete fitness function can be defined as \\
\begin{equation}
    J(r(t)) = V_c - \delta_{p_{\mathcal{O}}} P_0 - \delta_{p_{\mathcal{I}}} P_1 
    \label{eq:FitnessFunction}
\end{equation}\\
With the definition of the fitness function, relaxation of constraints and a re-sampled trajectory for a fixed time step $\Delta t$, the optimal control problem of maximising the measured gravitational signal can be simplified to solve the following continuous minimisation problem\\
\begin{equation}
    \begin{aligned}
        \min_{\boldsymbol{x}_0, \boldsymbol{u}} \quad & - V_{c} + \delta_{p_{\mathcal{O}}}P_{0} + \delta_{p_{\mathcal{I}}} P_{1}\\
        \textrm{s.t.} \quad & \boldsymbol{u} \in \mathcal{U}(\boldsymbol{x},t), \quad \forall t \in \mathcal{T}_{C}\\
    \end{aligned}
    \label{eq:OptimisationProblem}
\end{equation}\\
With the complex dynamics provided by $\dot{x}$ and the fact that the problem now deals with optimising numerous variables, such as the initial state $x_{0}$ and control strategies $\mathcal{U}(\boldsymbol{x}, t)$, it is highly motivated to adopt robust search processes that efficiently cover a large proportion of the defined search space. In this work, we will consider the evolutionary extended ant colony optimisation scheme as presented by Schl\"{u}ter et al. \cite{Schlter2009} and further implemented in PyGMO \cite{Pygmo}. To solve the problem, the chromosome is formulated as the vector\\
\begin{equation}
    \begin{aligned}
        \textit{chromosome } = & \text{ } [x_{0}, \mathcal{T}_{C}, \mathcal{U}(x, t \in \mathcal{T}_{C})] \\
        = & \text{ }[r_{0,x}, r_{0,y}, r_{0,z}, v_0, v_{0,x}, v_{0,y}, v_{0,z}, t_0, \Gamma(t_0), u_x (t_0), u_y (t_0), u_z (t_0), \dots, \\
        & \text{ } t_i, \Gamma(t_i), u_x (t_i), u_y (t_i), u_z (t_i), \dots, t_I, \Gamma(t_I), u_x (t_I), u_y (t_I), u_z (t_I)]
    \end{aligned}
\end{equation}\\
where $t \in \mathcal{T}_{C}$, $\Gamma_{min} \leq \Gamma(t_0) \leq \Gamma_{max}$, $v_{min} \leq v_0 \leq v_{max}$,  $r,v \in \mathbb{R}^3$ and $i = 0, \dots, I$. The initial velocity and sequence of control vectors are divided into two parts, a magnitude and a unit vector for direction, to increase the robustness and stability of the search process.

\subsection{Cooperative Coverage Optimisation}\label{section 3.2}
In contrast to Section \ref{section 3.1}, the problem is now expanded to consider multiple spacecraft. Formally, we introduce a new notation depicting a set of $K$ active spacecraft, $\mathcal{S}_{A} = \{s_1, \dots, s_k, \dots, s_K \}$. In this work, we consider the fitness function to compute the joint performance of the spacecraft. Visiting any specific cell on the spherical grid will only generate value for the fitness function once. Similarly, the risk of the corresponding swarm is related to the spacecraft closest to the body at any point on the considered trajectories. The far distance penalty will assume the total average deviation from the region of interest for all spacecraft. With these aspects in mind, the optimisation problem presented in Equation \ref{eq:OptimisationProblem} can be further expanded to cover the whole set $\mathcal{S}_{A}$  by introducing the extended chromosome\\
\begin{equation}
    \begin{aligned}
        \textit{chromosome }_{\textit{swarm}} = \text{ } & [x_{0}^{1}, \mathcal{T}_{C}^{1}, \mathcal{U}(x^{1}, t \in \mathcal{T}_{C}^{1}), \dots, \\
        & x_{0}^{k}, \mathcal{T}_{C}^{k}, \mathcal{U}(x^{k}, t \in \mathcal{T}_{C}^{k}), \dots, x_{0}^{K}, \mathcal{T}_{C}^{K}, \mathcal{U}(x^{K}, t \in \mathcal{T}_{C}^{K})]
    \end{aligned}
\end{equation}\\
for each spacecraft $k \in \mathcal{S}_{A}$.

%% file: Sections/Results.tex
\section{Results}\label{section 4}
\begin{figure}[h]
     \centering
     \begin{subfigure}[b]{0.32\textwidth}
         \centering
         \includegraphics[width=\textwidth]{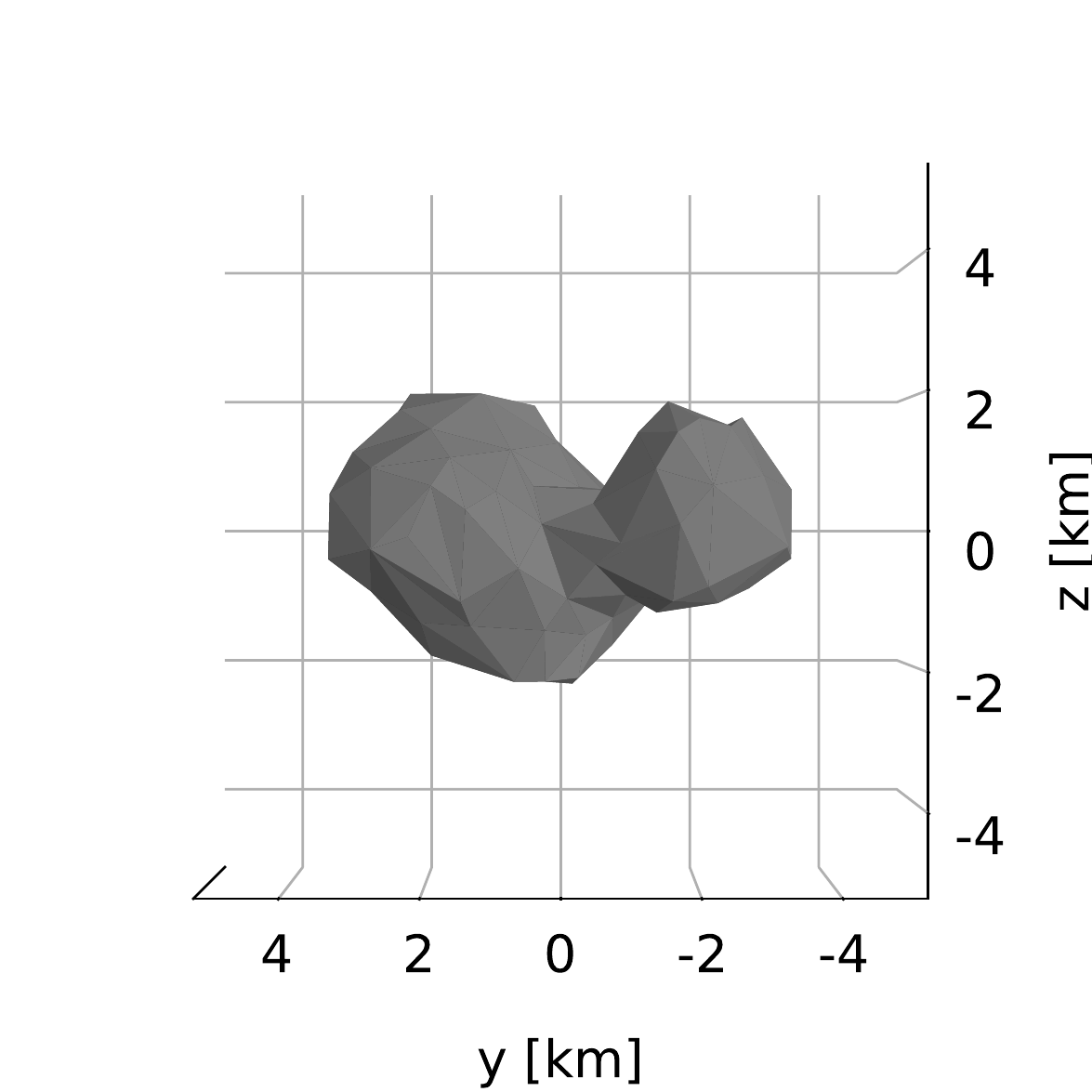}
         \caption{Low resolution mesh.}
         \label{fig: Low res mesh}
     \end{subfigure}
     \begin{subfigure}[b]{0.32\textwidth}
         \centering
         \includegraphics[width=\textwidth]{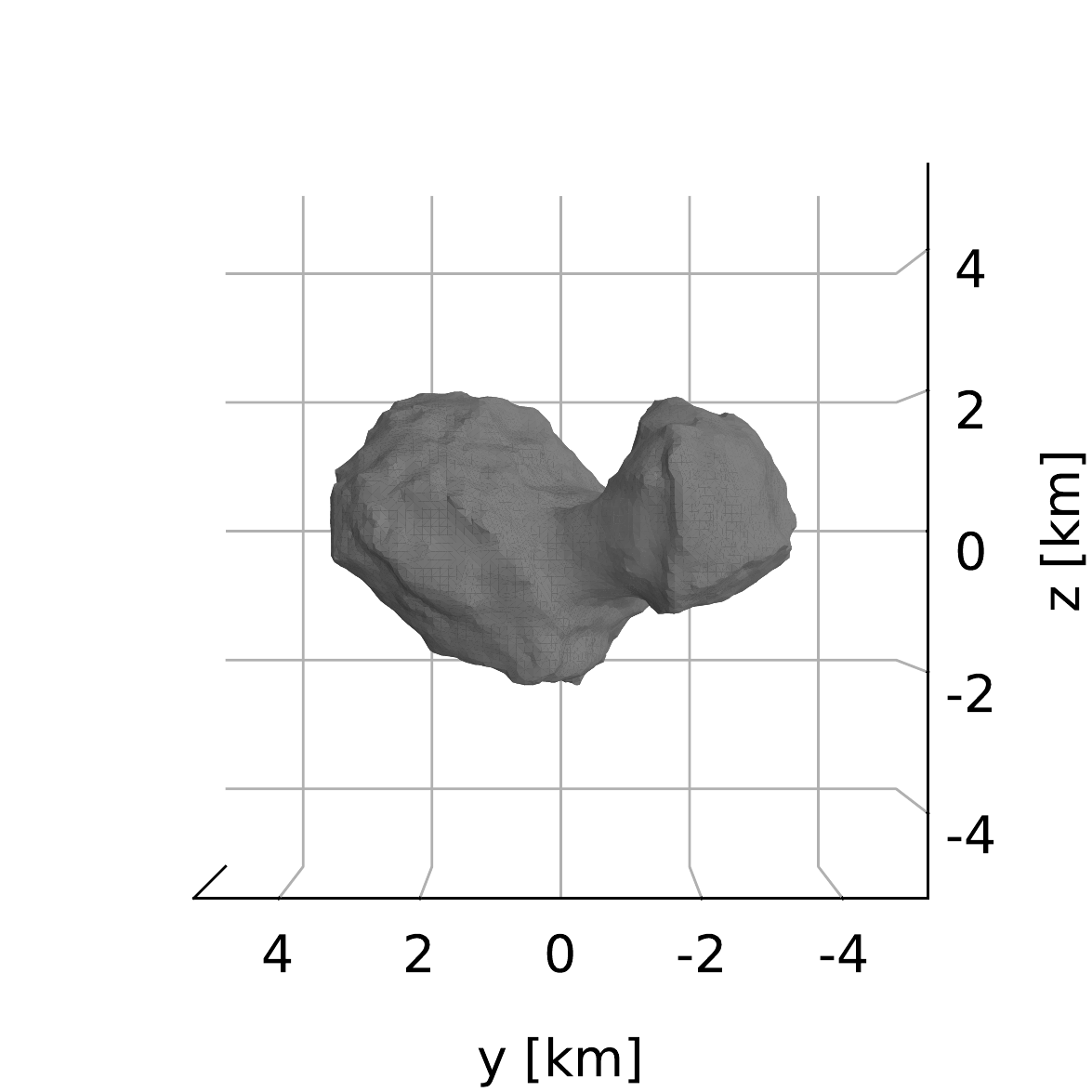}
         \caption{High resolution mesh.}
         \label{fig: High res mesh}
     \end{subfigure}
     \begin{subfigure}[b]{0.32\textwidth}
         \centering
         \includegraphics[width=\textwidth]{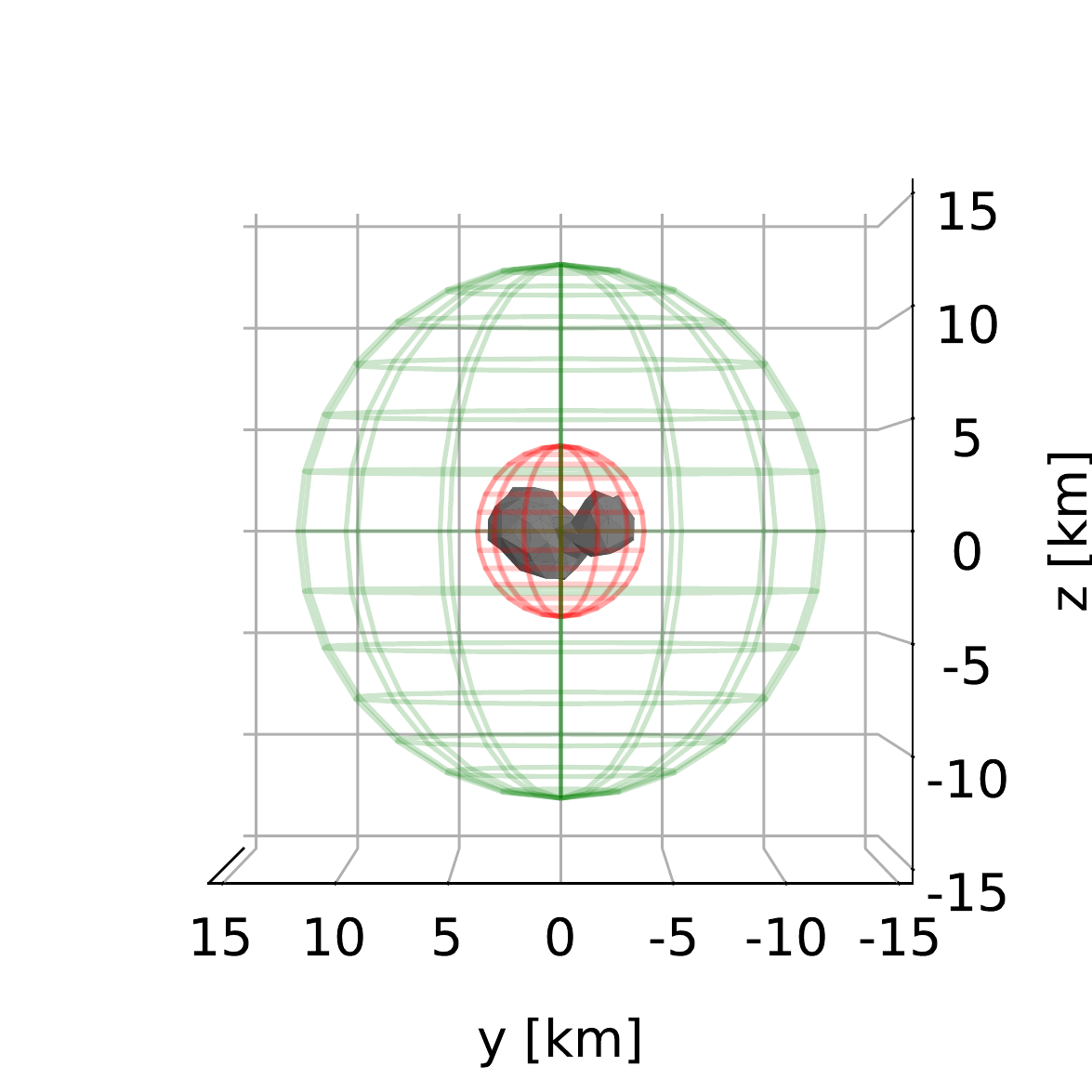}
         \caption{Bounding spheres.}
         \label{fig: Bounding spheres}
     \end{subfigure}
     \caption{Plot of comet depicting a) a low resolution mesh with $93$ vertices and $182$ faces}{ b) a high resolution mesh with $9149$ vertices and $18294$ faces and c) two bounding spheres.}
     \label{fig:mesh plots}
\end{figure}

In this section, we present the results of the proposed optimisation method. Specifically, we solve the optimisation problem for two test cases, the former using a single spacecraft and the latter a swarm consisting of four spacecraft. The results are reviewed for several performance measures, such as the evolution of measured signal quality through coverage, execution time, distance to the region of interest and fitness convergence rate. To consider a more realistic mission scenario, we use the comet 67P/Churyumov-Gerasimenko as the target small body since its complex shape allows for a challenging search process to define an optimal control law. As can be seen in Figure \ref{fig: High res mesh}, the comet consists of two lobes fused together by a narrow neck, where the large lobe is of dimensions $4.10 \times 3.52 \times 1.63$ km and the small lobe of dimensions $2.50 \times 2.14 \times 1.64$ km. Furthermore, the comet has a mass of $1.0\times10^{13}$ kg, a volume of $18.7$ $\text{km}^3$, a density of $533$ $\text{kg}/\text{m}^3$, a rotation period of $T_B = 12.06$ h and a principal axis with an orientation according to $69^{\circ}$ in right ascension and $64^{\circ}$ in declination \cite{ESARosetta}. Its ground-truth gravitational field is modelled by a polyhedral mesh consisting of $93$ vertices resulting in $182$ faces as presented in Figure \ref{fig: Low res mesh}. A low-poly version was used to minimize computational cost of the polyhedral model.

\subsection{Computational Setup}\label{section 4.1}
To evaluate the performance of the algorithm, we consider a large population pool of $120$ chromosomes and mutations for $1000$ generations. For each chromosome generated by the extended ant colony algorithm, the equations of motion are numerically integrated with an absolute and relative error tolerance of $1$e-$12$, respectively. Each candidate solution is then re-sampled for a fixed time step $\Delta t = 100$ s before evaluating the corresponding fitness function $J(r(t))$. The inner and outer bounding spheres, as depicted in Figure \ref{fig: Bounding spheres}, are defined by their fixed radius $r_{\mathcal{I}} = 4$ km and $r_{\mathcal{O}} = 12.5$ km, thus roughly $0.6$ km and, respectively, $9.1$ km above the comet's surface. Furthermore, each test case is simulated on two Intel(R) Xeon(R) Silver $4216$ CPUs at $2.10$ GHz with a total of $32$ cores and $64$ threads. The simulation is parallelised using $60$ threads, each evaluating one chromosome at a time.

\subsection{Single Spacecraft Trajectory Optimisation}\label{section 4.2}
\begin{figure}[h]
     \centering
     \begin{subfigure}[b]{0.32\textwidth}
         \centering
         \includegraphics[width=\textwidth]{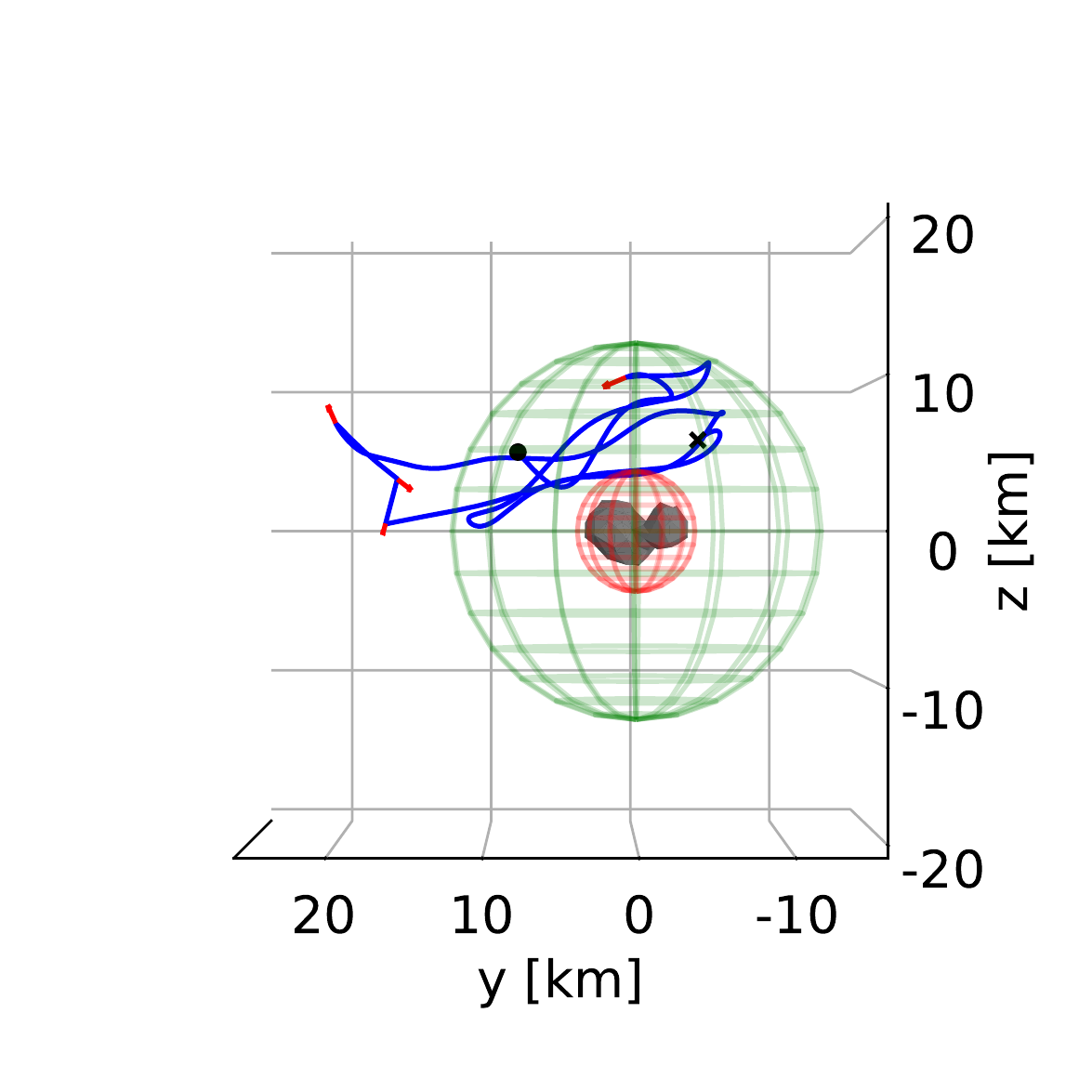}
         \caption{Side view.}
         \label{fig:SingleSpacecraftRotatingFrameA}
     \end{subfigure}
     \begin{subfigure}[b]{0.32\textwidth}
         \centering
         \includegraphics[width=\textwidth]{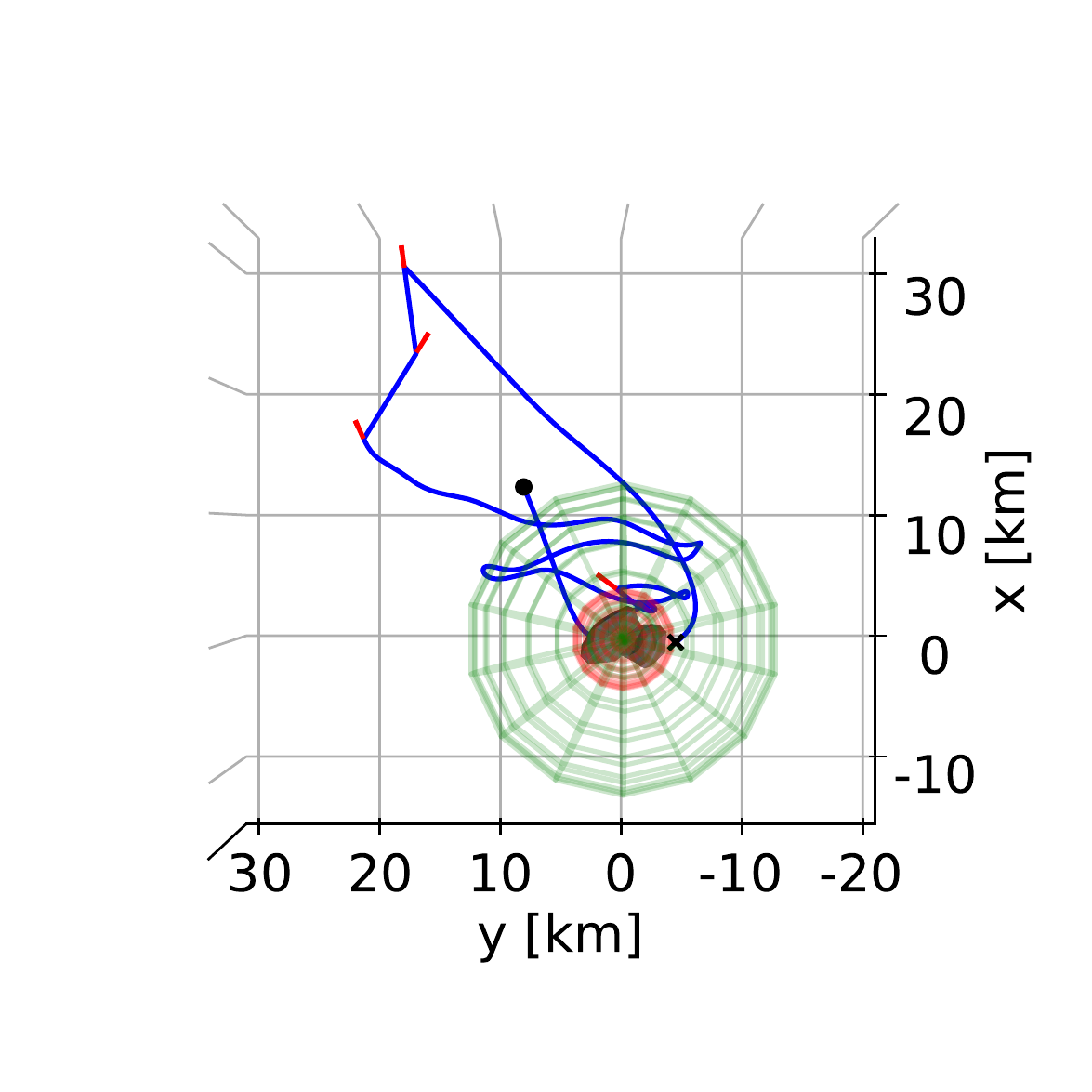}
         \caption{Top view.}
         \label{fig:SingleSpacecraftFixedBodyFrameB}
     \end{subfigure}
     \begin{subfigure}[b]{0.32\textwidth}
         \centering
         \includegraphics[width=\textwidth]{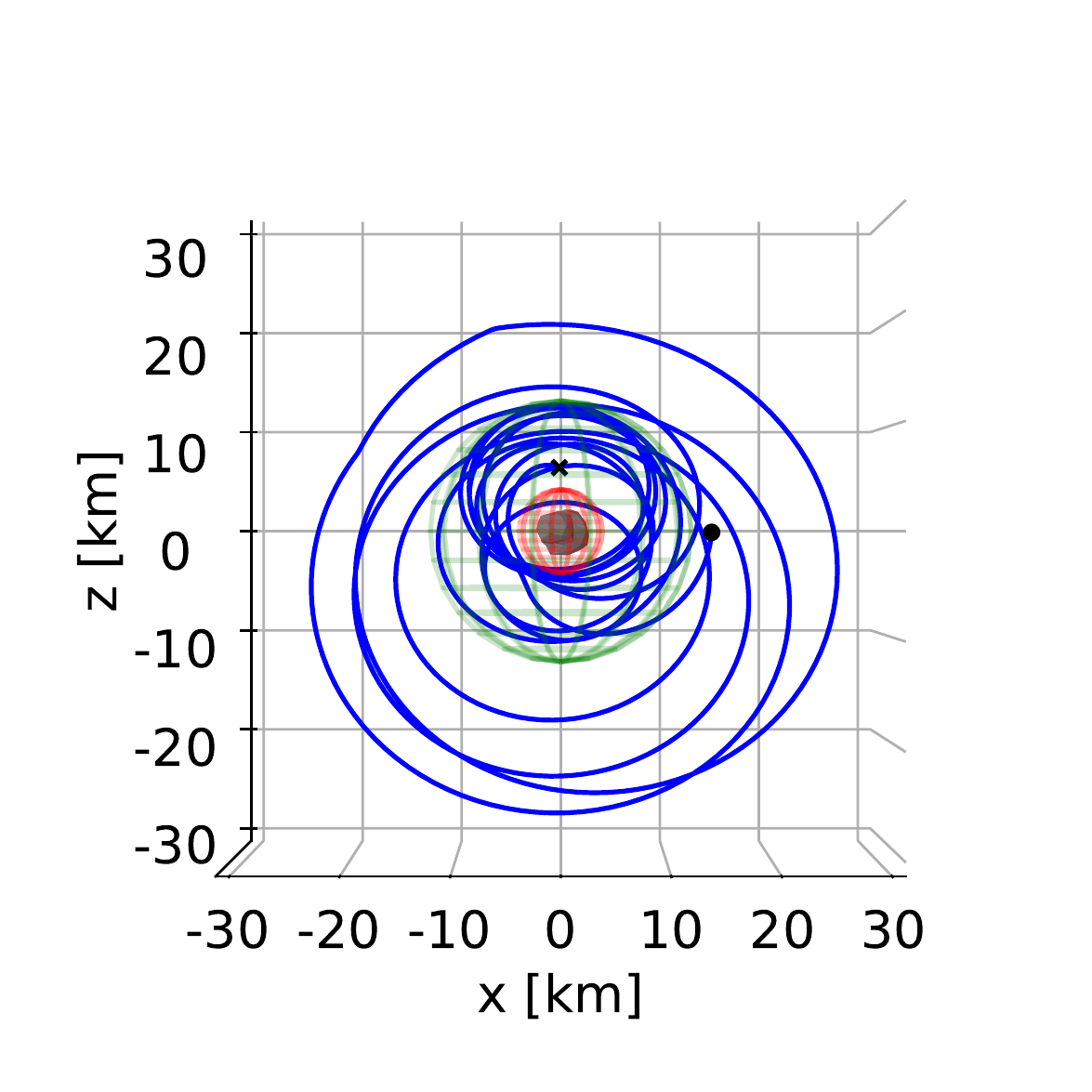}
         \caption{Body-fixed frame.}
         \label{fig:SingleSpacecraftFixedBodyFrame}
     \end{subfigure}
     \caption{Single spacecraft optimisation results depicting a) side view of the trajectory in the inertial}{ reference frame, b) corresponding top view and c) the resulting trajectory in the body-fixed frame.}
     \label{fig:SingleSpacecraftTrajectory}
\end{figure}

For the first test case, we consider a single spacecraft with four impulsive manoeuvres over a mission duration of seven days with an initial position fixed to $r_{0} = [-135, -4090, 6050]$ m. The initial velocity vector is optimised within a magnitude range of $v_{0} = [0,1.5]$ $\text{m}/\text{s}$ and unit direction $v_{0}$. Similarly, the control vectors have a defined magnitude range of $u_{t} = [0, 2.5]$ $\text{m}/\text{s}$ and unit direction $u_{t}$. The optimisation process finished in $262.4$ minutes and converged for a tolerance level of $1$e-$6$ after $280$ generations resulting in the trajectory presented in Figure \ref{fig:SingleSpacecraftTrajectory}. In Figure \ref{fig:SingleSpacecraftRotatingFrameA} and \ref{fig:SingleSpacecraftFixedBodyFrameB}, each manoeuvre is represented by a red arrow in the thrust direction. Immediately, it is noticeable that even though the spacecraft sometimes deviates from the region of interest, it tends to use its impulsive manoeuvres at its most distant positions to redirect its path towards the body. However, this behaviour is to be expected given the low gravitational field at these distances, resulting in particularly linear dynamics. Figure \ref{fig:SingleSpacecraftFixedBodyFrame} presents the corresponding trajectory in the body-fixed frame, which follows the expected circular motion around the body.

\begin{figure}[h]
     \centering
     \begin{subfigure}[b]{0.39\textwidth}
         \centering
         \includegraphics[width=\textwidth]{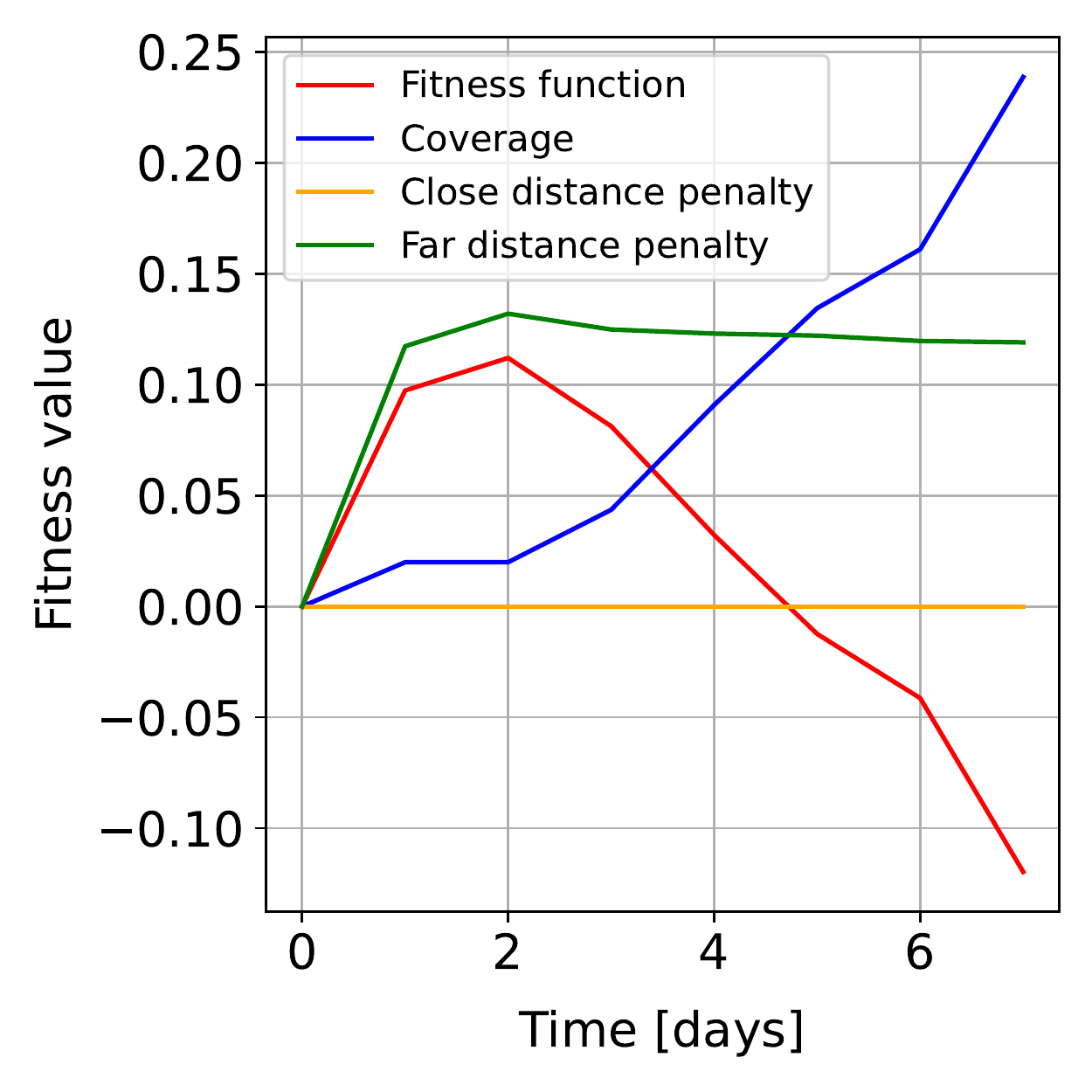}
         \caption{Evolution of fitness components.}
         \label{fig:SingleSpacecraftFitnessEvolution}
     \end{subfigure}
     \hspace{0.1\textwidth}
     \begin{subfigure}[b]{0.39\textwidth}
         \centering
         \includegraphics[width=\textwidth]{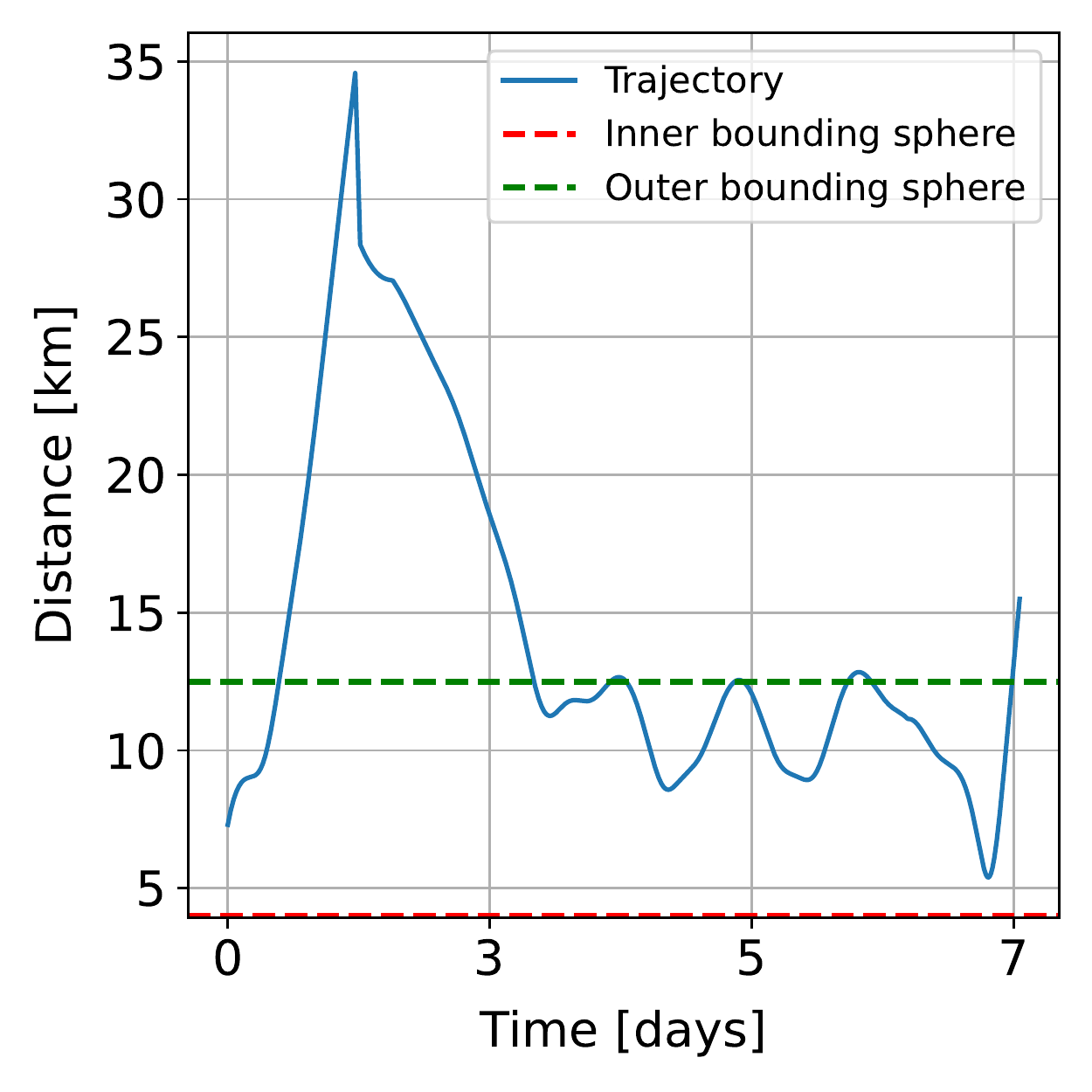}
         \caption{Distance from region of interest.}
         \label{fig:SingleSpacecraftDeviationOverTime}
     \end{subfigure}
     \caption{Single spacecraft optimization results depicting a) evolution of fitness components}{over the mission duration where the two penalties are minimised and the coverage is maximised, and b) distance from the center of the body.}
\end{figure}

Recall that the minimisation of the fitness function presented in Equation \ref{eq:FitnessFunction} maximises the coverage and minimises the two penalties. Figure \ref{fig:SingleSpacecraftFitnessEvolution} shows that the close distance penalty remains at zero, indicating that the spacecraft never enters the inner bounding sphere. The coverage increases almost monotonically except between days one and two, where it stagnates as the spacecraft deviates from the region of interest. This behaviour can also be seen in Figure \ref{fig:SingleSpacecraftDeviationOverTime}, which shows a resonating distance evolution to the comet's centre. Similarly, the far distance penalty stagnates at a moderately low value after three days representing that the spacecraft generally remains close to the body over the mission duration, which can be seen in Figure \ref{fig:SingleSpacecraftRotatingFrameA} and \ref{fig:SingleSpacecraftFixedBodyFrameB} as well.\\

\subsection{Multi-Spacecraft Trajectory Optimisation}\label{section 4.3}
\begin{figure}[h] 
    \centering
    \begin{subfigure}{0.32\textwidth}
        \centering
        \includegraphics[width=\linewidth]{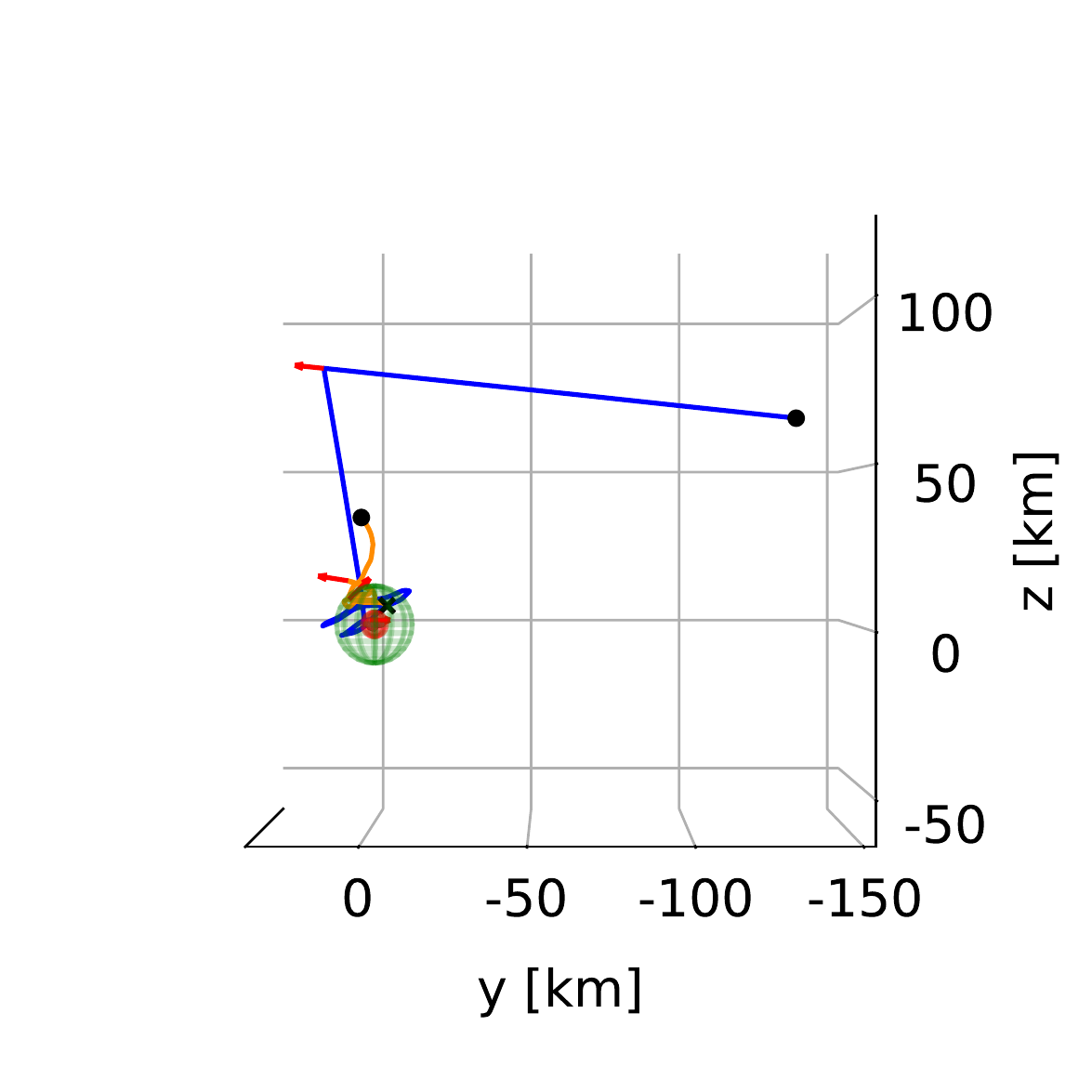}
        \caption{Spacecraft 1 \& 2 }{(side view).}
        \label{fig:a}
    \end{subfigure} 
    \begin{subfigure}{0.32\textwidth}
        \centering
        \includegraphics[width=\linewidth]{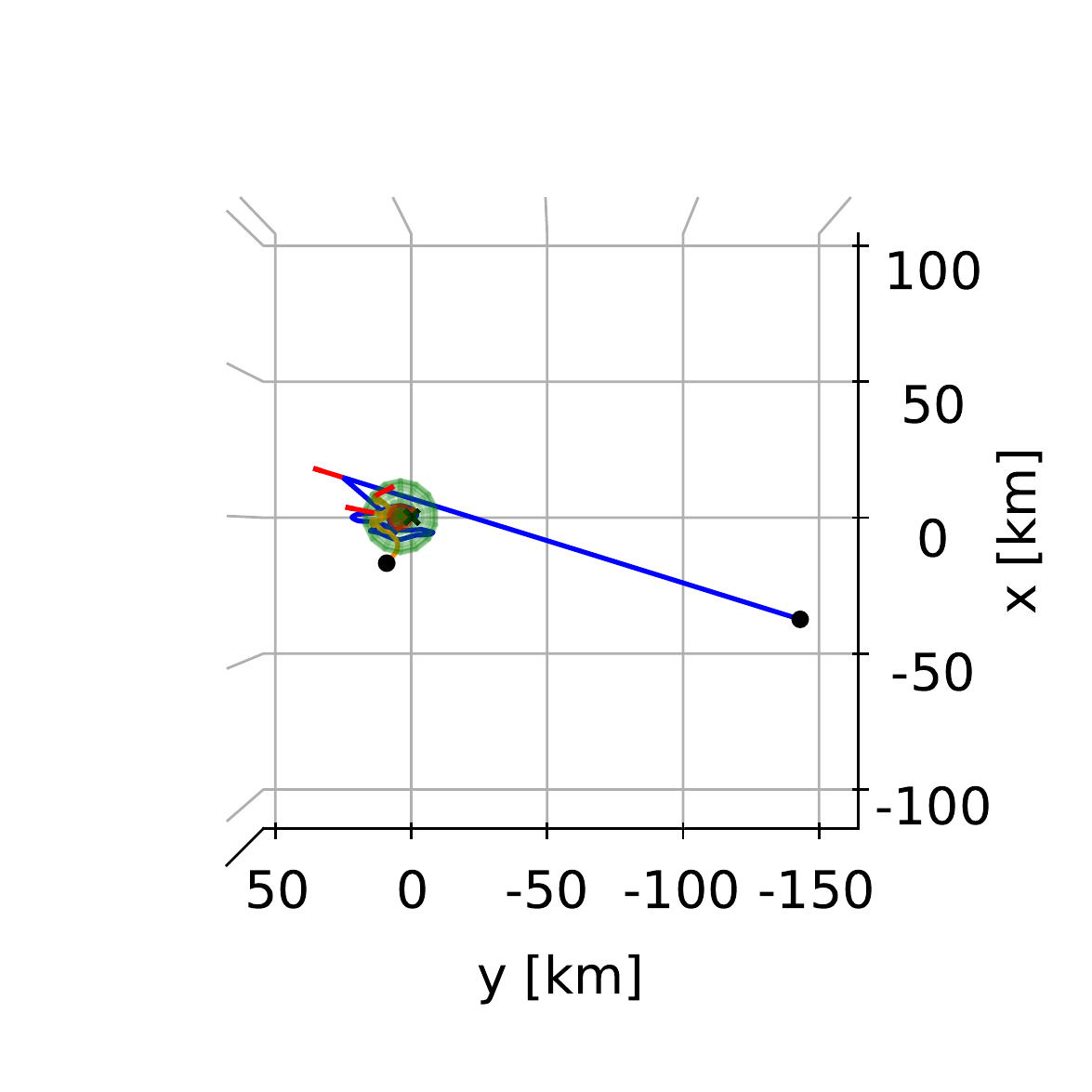}
        \caption{Spacecraft 1 \& 2 }{(top view).}
        \label{fig:b}
    \end{subfigure}
    \begin{subfigure}{0.32\textwidth}
        \centering
        \includegraphics[width=\linewidth]{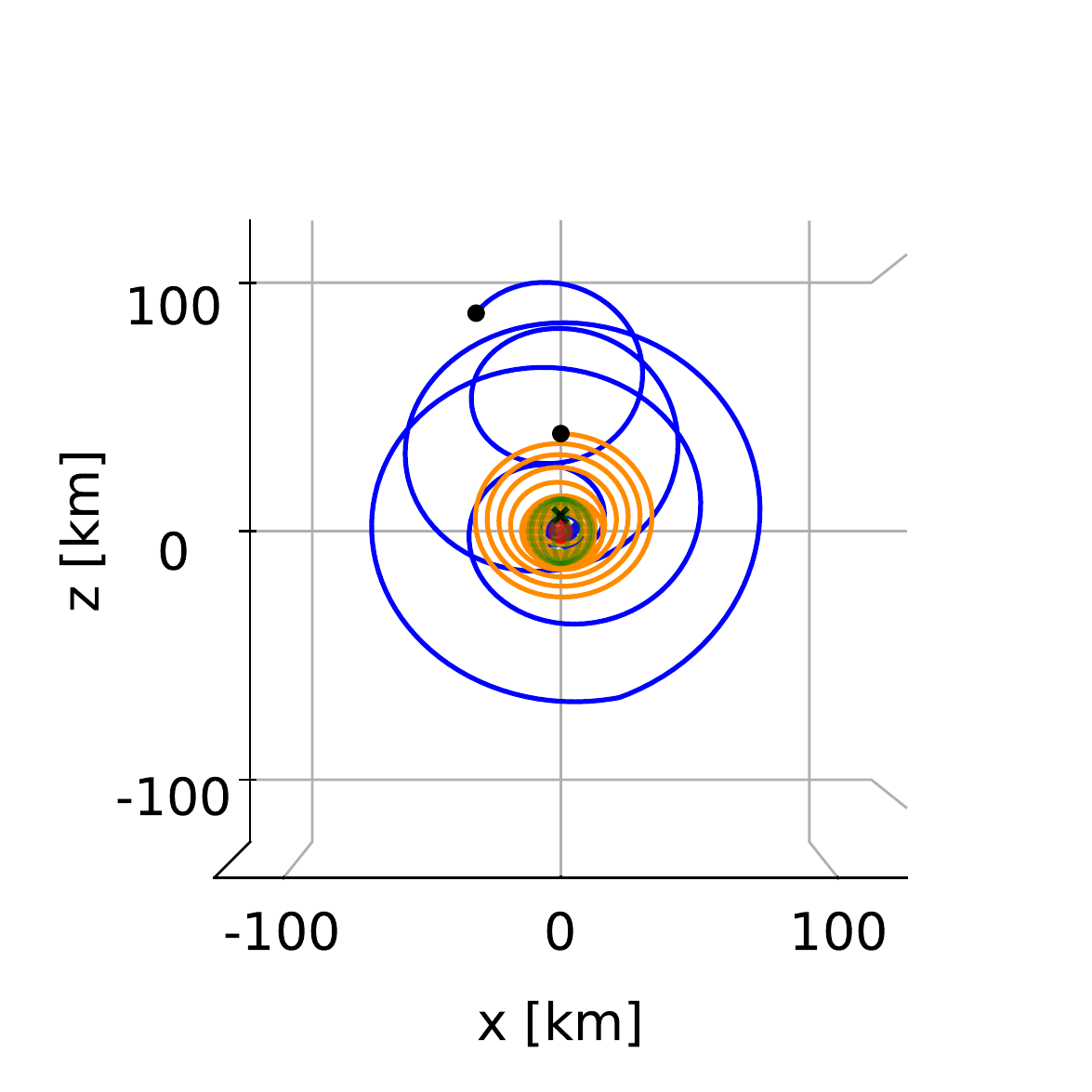}
        \caption{Spacecraft 1 \& 2 }{(body-fixed frame).}
        \label{fig:c}
    \end{subfigure}
    
    \medskip
    \begin{subfigure}{0.3\textwidth}
        \centering
        \includegraphics[width=\linewidth]{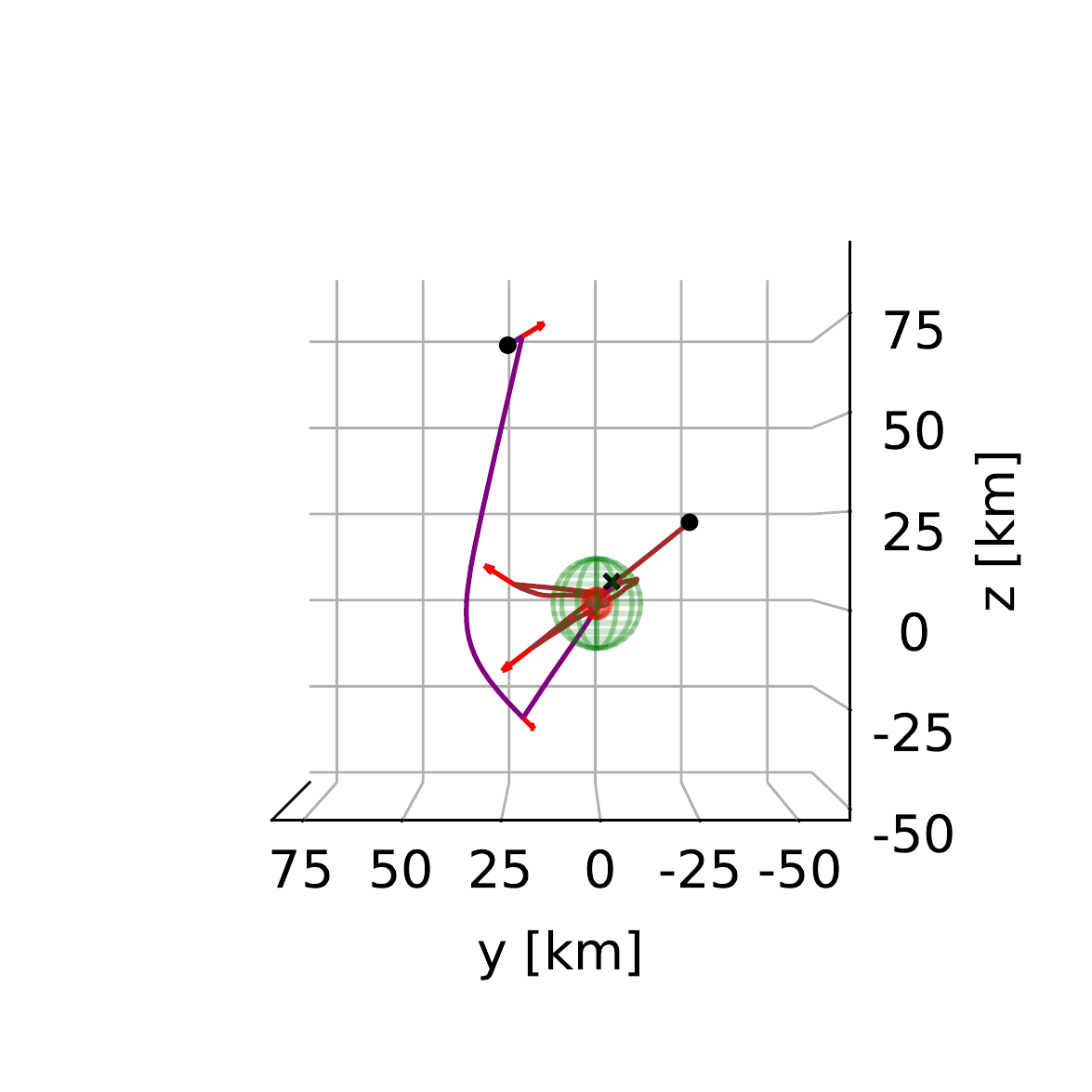}
        \caption{Spacecraft 3 \& 4 }{(side view).}
        \label{fig:d}
    \end{subfigure}
    \begin{subfigure}{0.3\textwidth}
        \centering
        \includegraphics[width=\linewidth]{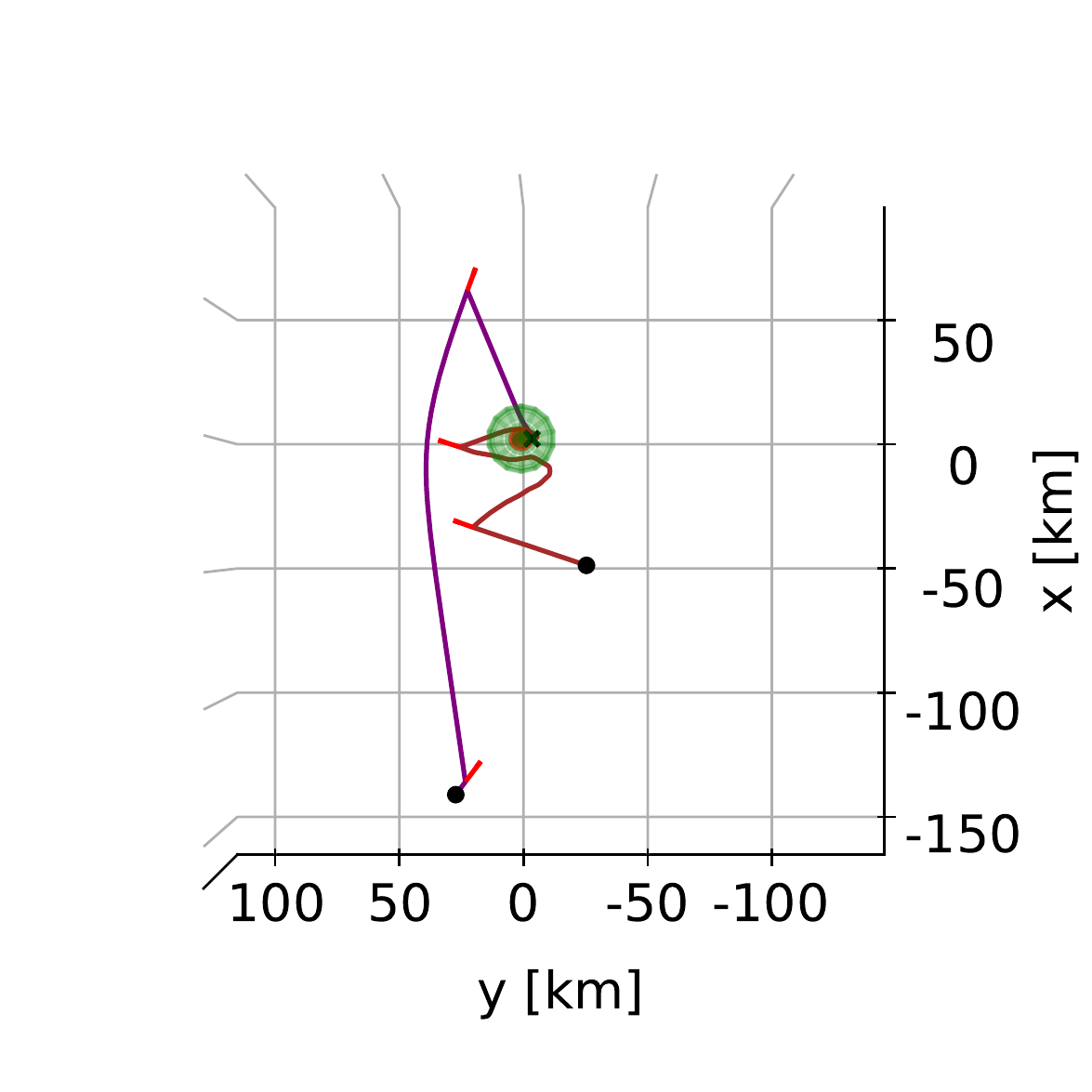}
        \caption{Spacecraft 3 \& 4 }{(top view).}
        \label{fig:e}
    \end{subfigure}
    \begin{subfigure}{0.3\textwidth}
        \centering
        \includegraphics[width=\linewidth]{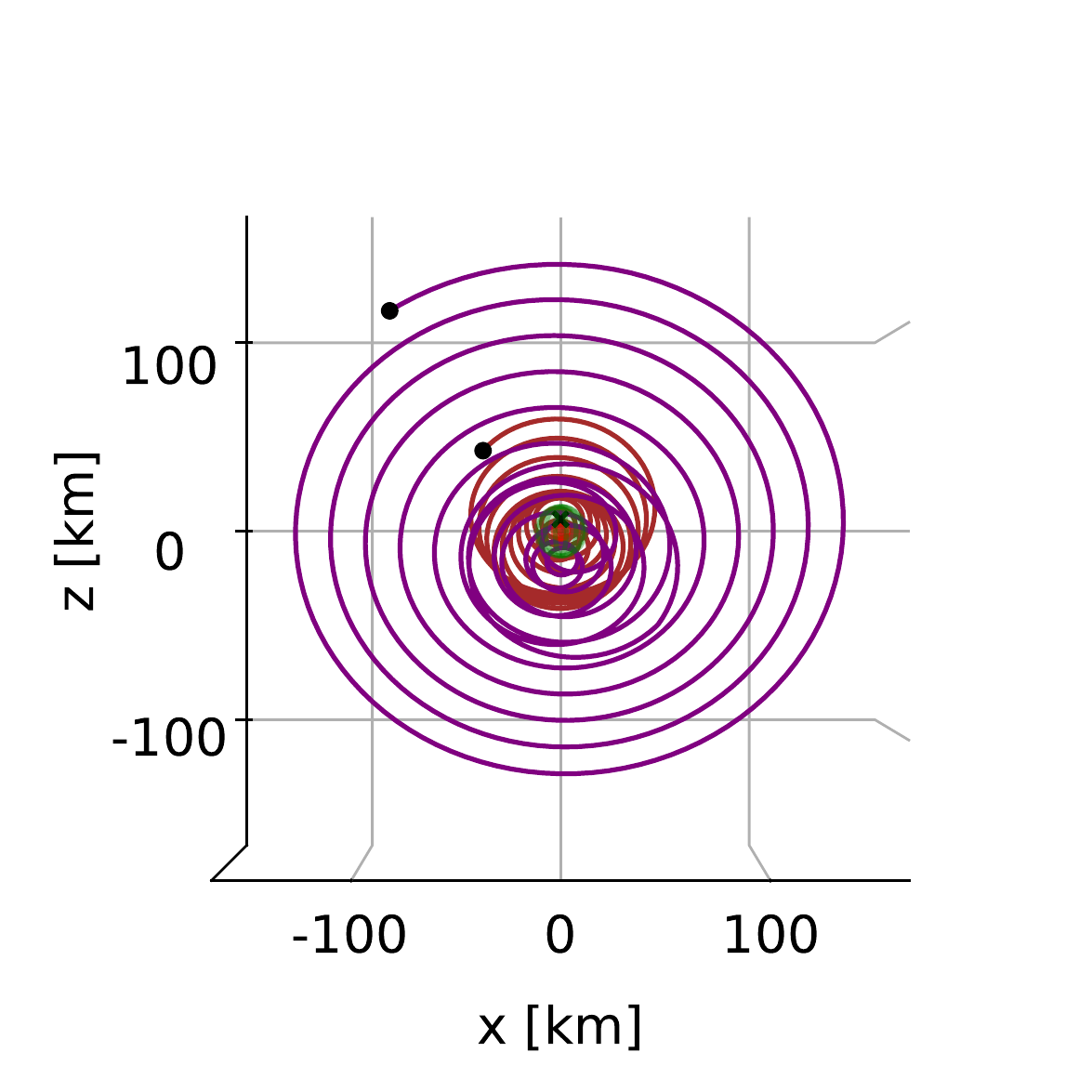}
        \caption{Spacecraft 3 \& 4 }{(body-fixed frame).}
        \label{fig:f}
    \end{subfigure}
    \caption{Multi-spacecraft optimisation results depicting a) - b) resulting trajectories for Spacecraft 1 and 2}{in the inertial reference frame, c) resulting trajectories in the body-fixed frame}{ and d) - f) similar representation for Spacecraft 3 and 4. For visibility and space reasons we show two spacecraft at a time.}
    \label{fig:MultiSpacecraftTrajectories}
\end{figure}

For the second test case, we consider a set of four spacecraft, each with two impulsive manoeuvres over the mission duration of seven days. The initial position is fixed to $r_{0} = [-135, -4090, 6050]$ m for each spacecraft, which could represent a deployment from a mothership similar to the Hera mission \cite{Michel2018}. The initial velocity vector is optimised within its allowed magnitude range of $v_{0} = [-1.5,1.5]$ $\text{m}/\text{s}$ and unit direction $v_{0}^{k}$. Similarly, the control vectors have a defined magnitude range of $u_{t} = [-2.5, 2.5]$ $\text{m}/\text{s}$ and unit direction $u_{t}^{k}$. The optimisation process finished after $850.92$ minutes and converged for an acceptable tolerance level of $1$e-$6$ after $297$ generations. In Figure \ref{fig:a} and \ref{fig:b}, Spacecraft $1$, as depicted in blue, travels generally at a greater distance from the body than Spacecraft $2$, in orange, which is contained within the region of interest for a longer period of the mission duration. Both spacecraft utilise their impulsive manoeuvre to return to the body from distant positions. In Figure \ref{fig:d} and \ref{fig:e}, Spacecraft $3$ (in brown) and $4$ (in purple) seem to follow a similar pattern, with Spacecraft $4$ travelling farthest away at approximately $170$ km from the center of the body. Furthermore, Figure \ref{fig:c} and \ref{fig:f} present the corresponding trajectories in the body-fixed frame which follow an expected circular motion according to the body's rotation.
\\

\begin{figure}[h!]
    \centering
    \begin{subfigure}[b]{0.39\textwidth}
        \centering
        \includegraphics[width=\textwidth]{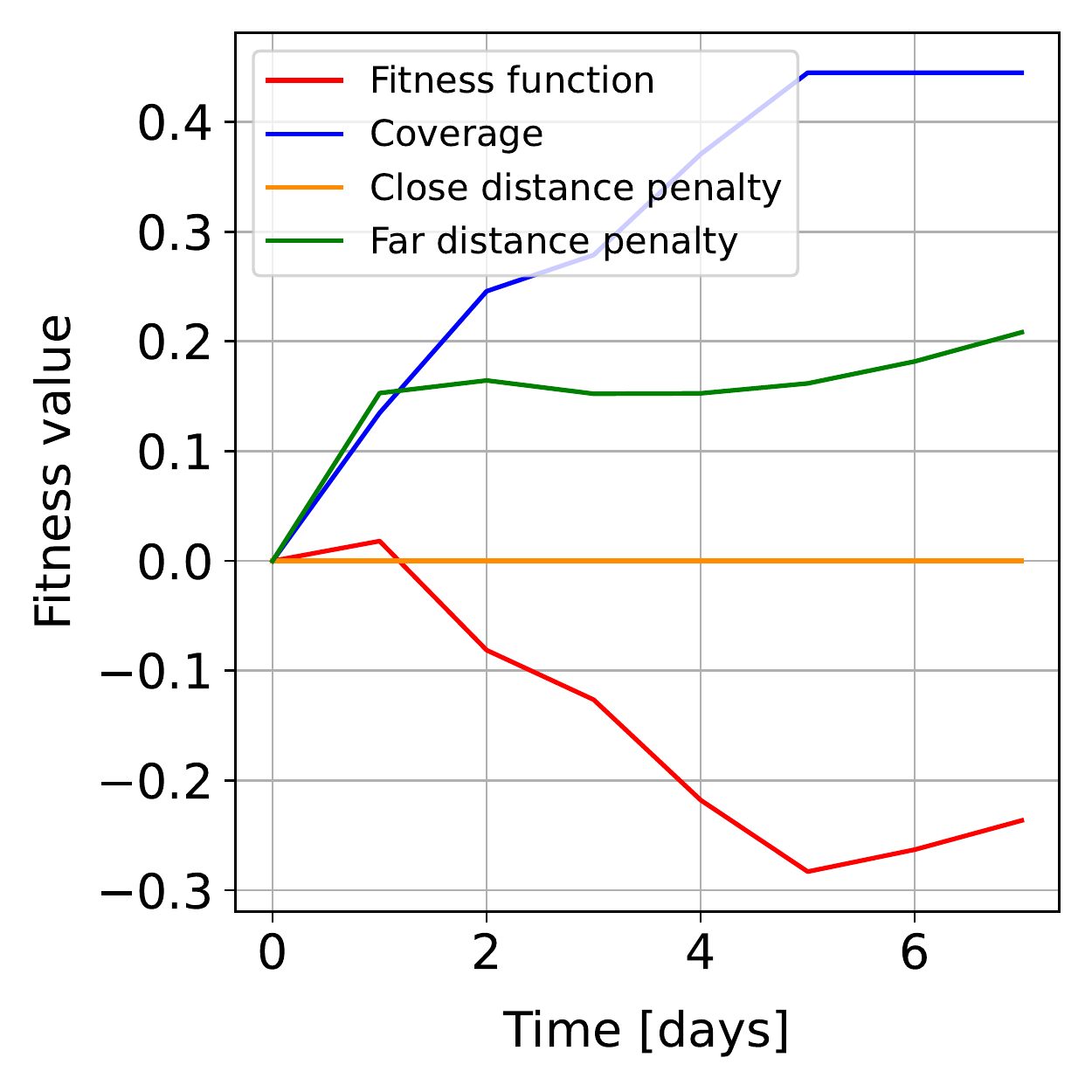}
        \caption{Evolution of fitness components.}
        \label{fig:MultiSpacecraftFitnessEvolution}
    \end{subfigure}
    \hspace{0.1\textwidth}
    \begin{subfigure}[b]{0.39\textwidth}
        \centering
        \includegraphics[width=\textwidth]{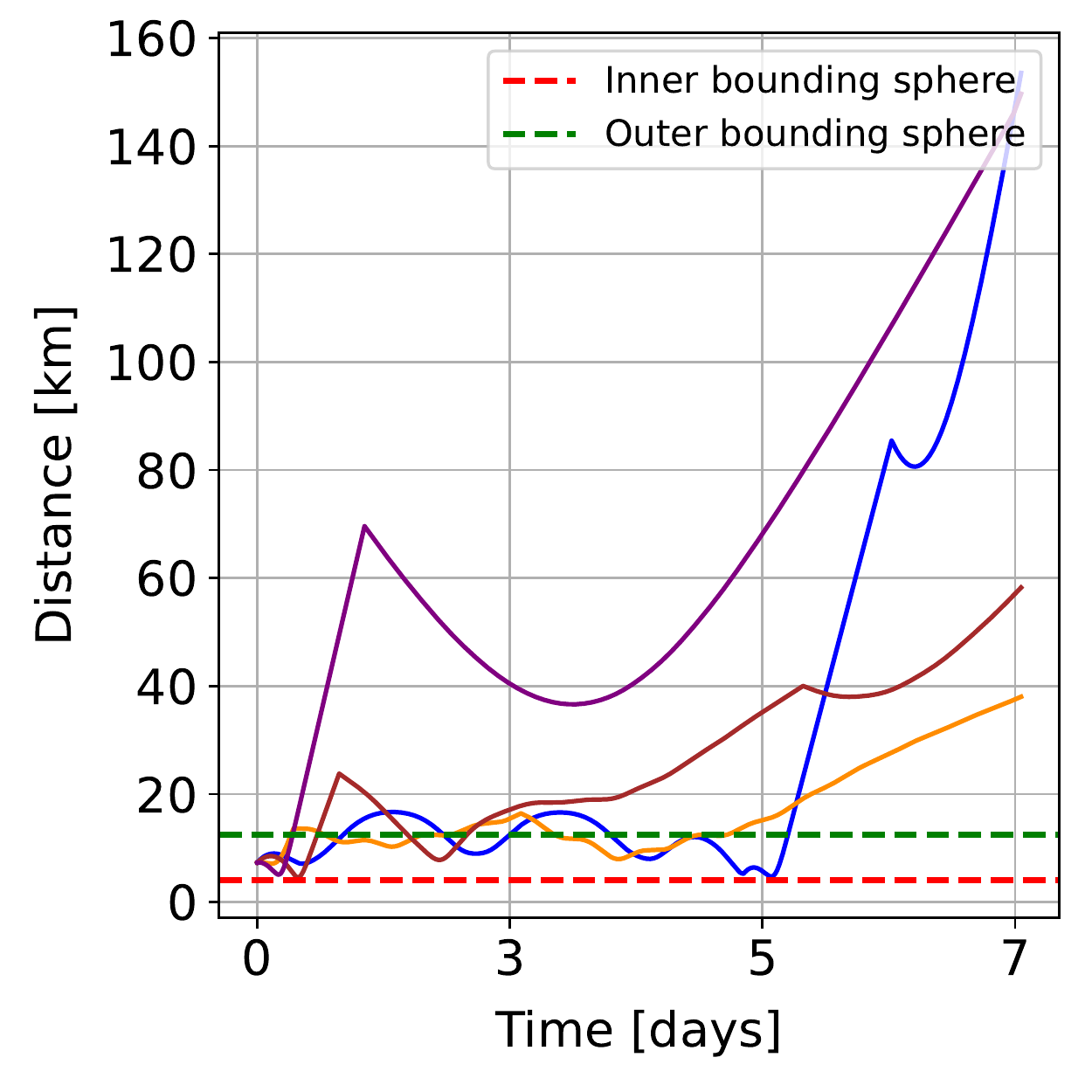}
        \caption{Distance from the region of interest.}
        \label{fig:MultiSpacecraftDeviationOverTime}
    \end{subfigure}
    \caption{Multi-spacecraft optimization results depicting a) evolution of fitness components}{over the mission duration where the two penalties are minimised and the coverage is maximised and b) distance from the center of the body.}
\end{figure}

From Figure \ref{fig:MultiSpacecraftFitnessEvolution}, it should be noted that the close distance penalty remains at zero, indicating the absence of positions inside the safety radius. The far distance penalty changes marginally after day one, and the coverage increases almost monotonically over the mission duration, except between days five and seven. The far distance penalty and the coverage are larger for the swarm than for a single spacecraft. Reviewing Figure \ref{fig:MultiSpacecraftDeviationOverTime}, it is evident that most spacecraft can maintain beneficial trajectories for the first three days, except Spacecraft $4$, which never returns to the region of interest after day one. In addition, it should also be noted that none of the considered spacecraft is able to successfully utilise its impulsive manoeuvres to revisit the region of interest after day five.

%% file: Sections/Discussion.tex
\section{Discussion}\label{section 5}
\subsection{Method Considerations}\label{section 5.1}
In this work, we have presented the formulation of a trajectory optimisation problem that, when solved efficiently, results in an initial state and optimal control sequence maximising the measured gravitational signal around an irregularly shaped rotating body using both a single and multiple spacecraft. By representing the region of interest with a spherical tensor grid, the coverage of a spacecraft can be approximated as the weighted ratio of visited tesseroids defined on the grid. On that note, however, the coverage is, in truth, of continuous nature and would subsequently be most accurately represented by the numerical integration of some continuous function. Although the proposed method provides other benefits, such as simplicity and reduced computational complexity, the trade-off between the discrete and continuous approximation should be considered when reviewing the results.

Furthermore, in order to avoid ruggedness and, subsequently, the potential for finding inefficient solutions, it is essential to scale the penalties to be comparable with the coverage. The penalties are scaled by assigning appropriate values for the corresponding Lagrangian multipliers, which in this work are referred to as penalty scaling factors. However, introducing additional hyperparameters also increases the optimisation complexity. Moreover, it should also be mentioned that the fitness function lacks parameters such as the sum of delta-v corresponding to the impulsive manoeuvres, thus enabling the optimisation to favour high-energy trajectories instead of more stable solutions.

Another delimiting factor of the proposed method is the number of variables considered for the decision vector. For instance, in the second case studying the application of four spacecraft with two possible manoeuvres each, the problem formulation resulted in a chromosome consisting of $56$ variables. Consequently, expanding the optimisation space makes it difficult for the extended ant colony algorithm to sample the state space adequately, resulting in possibly sub-optimal solutions. This behaviour becomes evident when studying the convergence rate for the two test cases.

\subsection{Computational properties}\label{section 5.2}
Reviewing the computational properties of the optimisation process, it is evident that propagating a trajectory using a polyhedral gravity model results in a computationally expensive task. In fact, evaluating positions using the model to obtain acceleration values for the state propagation makes up almost a third of the total simulation time. For comparison, the second and third most time-consuming tasks, being interpolation and quaternion rotations, make up for approximately $9\%$ and $5\%$ per cent of the total time, respectively. In order to reduce the time for solving the problem, one can instead use a polyhedral model with a reduced number of vertices resulting in a faster evaluation process. However, choosing the appropriate resolution of the mesh should be done considering the effects on accuracy for modelling the gravitational field \cite{GeodesyNetsBenchmark}. Apart from the polyhedral model, fine-tuning parameters such as the tolerance level for integrating the equations of motion and the kernel size for the ant colony optimiser could improve performance both computationally and qualitatively. The effects of these parameters should therefore be studied further in more extensive testing.

\subsection{Single Spacecraft vs. Swarm}\label{section 5.3}
For both test cases, it is evident that the algorithm converges for solutions that utilise manoeuvres to change the spacecraft's course rapidly, favouring larger control magnitudes at greater distances from the region of interest. The total delta-v used in the first test case is $2.779$ $\text{m}/\text{s}$, which is evenly distributed over its two impulsive manoeuvres and within a realistic magnitude range. For the swarm, Spacecraft $1$ used a total delta-v of $1.893$ $\text{m}/\text{s}$, Spacecraft $2$ used $0.186$ $\text{m}/\text{s}$, Spacecraft $3$ used $0.503$ $\text{m}/\text{s}$ and Spacecraft $4$ used $1.510$ $\text{m}/\text{s}$. Here, Spacecraft $2$ used the least amount of delta-v which is likely an effect of its proximity to the body, hence being able to utilise the gravitational acceleration and smaller correction manoeuvres to remain on a beneficial trajectory. However, in the case of Spacecraft $4$, which deviated to a larger extent from the body, it was necessary to use more delta-v in order to revisit the body. To motivate the use of more controlled path corrections, one could either increase the number of allowed manoeuvres or the corresponding magnitude range. However, increasing the number of manoeuvres also comes with the cost of increasing the search space, affecting the algorithm's convergence rate and ability to find a global optimum. For this reason, we evaluated several test cases with larger chromosomes than for the proposed swarm case, which generally resulted in less coverage, most likely due to the increasing complexity of the combinatorial problem.

Furthermore, comparing the optimisation performance of the two test cases, it is evident that the swarm was more effective than a single spacecraft, obtaining a $44\%$ coverage compared to $24\%$ over an equal mission duration. However, the individual performance of the single spacecraft in the first test case was greater than the individual coverage for the swarm. The behaviour is particularly evident in Figure \ref{fig:SingleSpacecraftTrajectory} and \ref{fig:SingleSpacecraftDeviationOverTime}, which shows a more stable trajectory for the first test case compared to Figure \ref{fig:MultiSpacecraftTrajectories} and \ref{fig:MultiSpacecraftDeviationOverTime} for the swarm. The results are likely an effect of the second case optimising four trajectories simultaneously in a high-dimensional scenario, thus leading to rapid growth in complexity. To avoid the challenge of dimensionality, a strategy for improving the swarm's performance could be to solve multiple single spacecraft trajectory optimisation problems and then combine the most appropriate solutions into an optimal set of trajectories. Apart from evaluating the joint coverage, the constellation's performance could also consider measures such as inter-spacecraft line of sight and total delta-v.

%% file: Sections/Conclusion.tex
\section{Conclusion}\label{section 6}
In conclusion, this work has studied the possibility of using an evolutionary optimisation algorithm for defining optimal trajectories of single and multiple spacecraft swarms to maximise the measured gravitational signal around a small body with challenging dynamics. The results indicate that the spacecraft swarm can effectively achieve better coverage than a single spacecraft over an equal mission duration, which is to be expected. However, optimising the trajectory for a single spacecraft compared to a swarm reduces the search space significantly, leading to a more efficient solution. It is therefore motivated to study how the dimensionality can be reduced for more complex scenarios to generate more efficient swarm solutions. Future work will consider adding several perturbing forces, such as the effects of solar radiation pressure, low-thrust manoeuvres and line of sight, including additional tools to model operational constraints \cite{PASEOS}. Another consideration is to benchmark how the resulting quality of the measured signal varies with the accuracy of the corresponding coverage function by comparing the spherical tensor with the numerical integration of a continuous model. Finally, the presented code base was made with modularity in mind, thus enabling an application to a broader range of scientific objectives to evaluate the performance of spacecraft swarms.

%% file: Sections/Acknowledgements.tex
\section{Acknowledgements}\label{section 7}
The authors would like to thank Christer Fuglesang and KTH Space Center for financially supporting the presentation of this work at the ICATT 2023 conference. 

Furthermore, the authors would also like to thank Filip Berendt and Ingemar Markström from The Visualization Studio, VIC, at KTH for providing an illustration of the spherical tensor grid. 